\documentclass[aps,twocolumn]{revtex4-2}
\usepackage[utf8]{inputenc}
\usepackage{microtype}
\usepackage{newtxtext}
\usepackage[upint]{newtxmath}
\usepackage{eucal}

\usepackage{graphicx}
\usepackage{booktabs}
\usepackage{enumerate}
\usepackage{siunitx}
\usepackage{braket}
\usepackage{color}
\usepackage{float}
\usepackage{soul}
\usepackage{bm}

\usepackage[colorlinks,allcolors=blue]{hyperref}
\usepackage[capitalize]{cleveref}

\usepackage[svgnames]{xcolor}
\usepackage{todonotes}
\setuptodonotes{inline}

\usepackage{comment}
\usepackage{lipsum}
\usepackage{float}

\usepackage[colorlinks,allcolors=blue]{hyperref}
\usepackage[capitalize]{cleveref}

\usepackage{braket}
\usepackage{tikz}
\usetikzlibrary{calc,patterns,angles,quotes}
\usetikzlibrary{angles} 

\definecolor{DarkRed}{rgb}{0.65,0,0}%
\definecolor{Green}{rgb}{0,0.3,0.3}
\definecolor{Purple}{rgb}{0.3,0,0.65}
\definecolor{Red}{rgb}{1,0,0}
\definecolor{Blue}{rgb}{0,0,0.85}
\definecolor{Magenta}{rgb}{1,0,1}

\newcommand{\ve}[1]{\boldsymbol{#1}}

\newcommand{\vectau}{\boldsymbol{\tau}}

\newcommand{\e}[1]{\mathrm{e}^{#1}}

\newcommand{\vecsigma}{\boldsymbol{\sigma}}

\newcommand{\vecd}{\ve{d}}

\newcommand{\vecM}{\ve{M}}

\newcommand{\eg}{\textit{e.g. }}

\def\i{\mathrm{i}}

\newcommand{\be}{\begin{equation}}
\newcommand{\ee}{\end{equation}}

\newcommand{\prlsection}[1]{\textit{#1}.\kern0.05em---\kern0.05em\ignorespaces}

\newcommand{\gR}[0]{g_{R}}
\newcommand{\tgR}[0]{\Tilde{g}_R}
\newcommand{\fR}[0]{f_{R}}
\newcommand{\tfR}[0]{\Tilde{f}_R}
\newcommand{\dgR}[0]{ g_R'}
\newcommand{\dtgR}[0]{\tgR'}
\newcommand{\dfR}[0]{ \fR'}
\newcommand{\dtfR}[0]{ \tfR'}
\newcommand{\dgL}[0]{ g_L'}
\newcommand{\dtgL}[0]{\tgL'}
\newcommand{\dfL}[0]{ \fL'}
\newcommand{\dtfL}[0]{ \tfL'}
\newcommand{\NR}[0]{N_R}
\newcommand{\tNR}[0]{\Tilde{N}_R}
\newcommand{\gmR}[0]{\gamma_R}
\newcommand{\tgmR}[0]{\tilde{\gamma}_R}
\newcommand{\sigmaxc}[0]{\sigmax^{*}}
\newcommand{\sigmayc}[0]{\sigmay^{*}}
\newcommand{\sigmaz}[0]{\tau_z}
\newcommand{\sigmazc}[0]{\sigmaz^{*}}
\newcommand{\sigmavec}[0]{\vecb{\tau}}
\newcommand{\gm}[0]{\gamma}
\newcommand{\tgm}[0]{\Tilde{\gamma}}

\newcommand{\tdxL}[0]{\Tilde{d}_{x,L}}
\newcommand{\tfs}[0]{\tilde{f}_s}
\newcommand{\tdx}[0]{\tilde{d}_x}
\newcommand{\tdy}[0]{\tilde{d}_y}
\newcommand{\tdz}[0]{\Tilde{d}_z}
\newcommand{\Uhat}[0]{\hat{U}}
\newcommand{\Ubar}[0]{\hat{U}}
\renewcommand{\i}{i}
\newcommand{\vecb}[1]{\boldsymbol{#1}}
\newcommand{\ex}[1]{\text{e}^{#1} }
\newcommand{\rhotre}[0]{\hat{\rho}_3}
\newcommand{\gbar}[0]{\underline{g}}
\newcommand{\tgbar}[0]{\underline{\tilde{g}}}
\newcommand{\fbar}[0]{\underline{f}}
\newcommand{\tfbar}[0]{\underline{\Tilde{f}}}
\newcommand{\tgamma}[0]{\Tilde{\gamma}}
\newcommand{\tN}[0]{\Tilde{N}}
\newcommand{\sgn}[0]{\, \text{sgn}}
\newcommand{\com}[2]{[#1,#2]}
\newcommand{\anticom}[2]{ \{#1, #2 \} }
\newcommand{\mhat}[0]{\hat{m}}
\newcommand{\gLhat}[0]{\hat{g}_L}
\newcommand{\gRhat}[0]{\hat{g}_R}
\newcommand{\ghat}[0]{\hat{g}}
\newcommand{\sigmaparvec}[0]{\hat{\vecb{\tau}}_{\|}}
\newcommand{\sigmaparx}[0]{\hat{\tau}_{\|,x}}
\newcommand{\dx}[0]{d_x}
\newcommand{\dy}[0]{d_y}
\newcommand{\dz}[0]{d_z}
\newcommand{\fs}[0]{f_s}
\newcommand{\dd}[0]{\text{d}}

\renewcommand{\Im}{\mathrm{Im}}
\renewcommand{\Re}{\mathrm{Re}}
\newcommand{\psidag}[0]{\psi^{\dagger}}
\newcommand{\Tr}[0]{\mathrm{Tr}}
\newcommand{\delz}[0]{\partial_z}
\newcommand{\dely}{\partial_y}
\newcommand{\fhat}{\hat{f}}

\newcommand{\dzR}[0]{d_{z,R}} 
 
\newcommand{\dxR}[0]{d_{x,R}} 
\newcommand{\tdxR}[0]{\tilde{d}_{x,R}} 
\newcommand{\dyR}[0]{d_{y,R}} 
 
\newcommand{\fsR}[0]{f_{s,R}}
\newcommand{\tfsR}[0]{\tilde{f}_{s,R}}
\newcommand{\delzfsL}[0]{\delz \fsL}
\newcommand{\delzdxL}[0]{\delz \dxL}
\newcommand{\prefen}[0]{\tfrac{2}{3} \tfrac{T_1^{2} p_{F}^{2}}{D}}
\newcommand{\prefto}[0]{mT_0 T_1}
\newcommand{\preftre}[0]{ d \alpha}
\newcommand{\fsL}[0]{f_{s,L}}
\newcommand{\dxL}[0]{d_{x,L}}
\newcommand{\dyL}[0]{d_{y,L}}
\newcommand{\dzL}[0]{d_{z,L}}
\newcommand{\tfsL}[0]{\tilde{f}_{s,L}}
\newcommand{\tdyL}[0]{\tilde{d}_{y,L}}
\newcommand{\prefnull}[0]{\tfrac{T_0^2}{D}}

\begin{document}
\title{Converting a triplet Cooper pair supercurrent into a spin-signal}
\author{Sigrid Aunsmo}
\affiliation{Center for Quantum Spintronics, Department of Physics, Norwegian \\ University of Science and Technology, NO-7491 Trondheim, Norway}
\author{Jacob Linder}
\affiliation{Center for Quantum Spintronics, Department of Physics, Norwegian \\ University of Science and Technology, NO-7491 Trondheim, Norway}

\begin{abstract}
Superconductivity with spin-polarized Cooper pairs is known to emerge by combining conventional spinless superconductors with materials that have spin-dependent interactions, such as magnetism and spin-orbit coupling. This enables a  dissipationless and conserved flow of spin. However, actually utilizing the spin-polarization of such supercurrents have proven challenging. Here, we predict an experimental signature of current-carrying triplet Cooper pairs in the form of an induced spin-signal. We show that a supercurrent carried only by triplet Cooper pairs induces a non-local magnetization that is controlled by the polarization direction of the triplet Cooper pairs. This provides a measurement protocol to directly use the spin-polarization of the triplet Cooper pairs in supercurrents to transfer spin information in a dissipationless manner.
\end{abstract}
\maketitle

\section{Introduction.} 

Substantial efforts have been made to find evidence of dissipationless currents carried by spin-polarized Cooper pairs in superconductors \cite{eschrig_njp_15, linder_nphys_15}, also referred to as triplet Cooper pairs. This includes measurements of how Gilbert damping is renormalized in Josephson junctions and superconducting bilayers subject to ferromagnetic resonance \cite{bell_prl_08, jeon_natmat_18, jeon_prb_19, yao_sciadv_21, muller_prl_21, carreira_prb_21}, as well as long-ranged supercurrent flow through magnetic materials \cite{keizer_nature_06, robinson_science_10, khaire_prl_10}. Spin-pumping in ferromagnet/superconductor (F/S) hybrid structures \cite{houzet_prl_08, yokoyama_prb_09}, where triplet pairing will affect the spin accumulation in the superconductor, has recently been subject of increased theoretical interest \cite{kato_prb_19, silaev_prb_20, simensen_prb_21, ahari_prb_21, ominato_prb_22, ojajarvi_prl_22, sun_prb_23, montiel_prb_23}.

However, a long-ranged supercurrent through magnetic materials is in itself not necessary useful. Supercurrents through normal metals carried by spinless, singlet Cooper pairs are also long-ranged when flowing through a normal metal. Thus, to unlock the potential of spin-polarized supercurrents with regard to potential cryogenic devices, it is necessary to find a way to utilize their spin-polarization directly. We here predict that a supercurrent carried by spin-polarized Cooper pairs induces a non-local magnetization. Both the polarization direction and magnitude of this magnetization is directly controlled by the spin degree of freedom of the triplet Cooper pairs. This shows how spin supercurrents can be used for low-dissipation information transfer by inducing spin-signals.

\begin{figure}[t!]
\includegraphics[width=0.87\columnwidth]{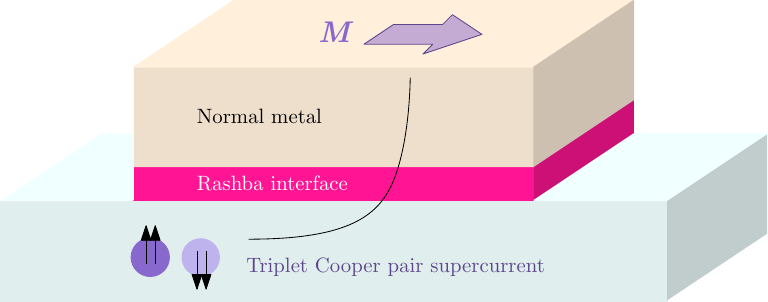}
	\caption{(Color online) Sketch of the proposed experimental protocol for converting a supercurrent carried by spin-polarized triplet Cooper pairs into a non-local spin signal. The supercurrent flows in a region underneath a thin heavy metal layer. Due to interfacial Rashba spin-orbit coupling, the triplet pairs induce a magnetization in a normal metal where no supercurrent is flowing. The induced magnetization depends sensitively on the polarization direction of the triplet pairs. 
	}
	\label{fig:model}
\end{figure}

The supercurrent flow is converted into a non-local magnetization, occuring in a region where the supercurrent does not flow, by allowing it to interact with a Rashba spin-orbit coupled interface. This can be probed in the setup shown in Fig. \ref{fig:model}. We consider the scenario experimentally realized in \eg Refs. \cite{khaire_prl_10, robinson_science_10}: a spin-triplet charge supercurrent generated by a magnetic multilayered structure. In the experiments, the spin-polarized nature of the Cooper pairs carrying the current was not directly measured, but rather inferred from its slow decay as a function of the length of the ferromagnetic bridge in a Josephson junction. In contrast, we provide a way to directly convert the spin of the triplet supercurrent into a spin-signal. The induced magnetization $\vecM$ in the normal metal (N) changes direction depending on the spin-polarization of triplet pairs carrying the supercurrent. Without current, the non-local magnetization vanishes. The induced magnetization $\vecM$ vanishes for certain spin-polarization directions of the pairs relative the Rashba interface normal. The predicted effect provides a way to directly use the spin-polarization of the triplet Cooper pairs in supercurrents to transfer spin information in a dissipationless manner.

\section{Theory} 
The quasiclassical theory of superconductivity \cite{usadel_prl_70, serene_physrep_83, eilenberger_zphys_68} is documented to compare well with experimental results, even quantitatively, for measurements done in mesoscopic superconducting hybrid structures. Our starting point is the Usadel equation which can be used in the diffusive limit of transport where the length scale hierarchy $\lambda_F \ll l_\text{mfp} \ll \xi$  applies, with $\lambda_F$ being the Fermi wavelength, $l_\text{mfp}$ the electronic mean free path, and $\xi$ the superconducting coherence length. It is effectively an equation of motion for the Green function matrix $\check{g}$ in Keldysh space and takes the form
\begin{align}
    D \nabla (\check{g} \nabla \check{g}) + \i[E\hat{\rho}_3 + \hat{\boldsymbol{h}} + \hat{\Delta}, \check{g}] = 0.
\end{align}
Here, $D$ is the diffusion coefficient, $E$ is the quasiparticle energy, $\hat{\boldsymbol{h}} = \text{diag}(\boldsymbol{h} \cdot \vectau, \boldsymbol{h} \cdot \vectau^*)$ where $\boldsymbol{h}$ describes the spin-splitting field, $\vectau$ is a vector with Pauli matrices as its components, $\hat{\rho}_3=\text{diag}(1,1,-1,-1)$, and $\hat{\Delta}$ describes the superconducting order parameter. The quasiclassical Green function $\check{g}$ is an 8$\times$8 matrix in Keldysh space and thus has a retarded, advanced, and Keldysh component \cite{rammer_rmp_86}. Since we will consider a system in equilibrium, $\check{g}$ is entirely determined by the retarded Green function $\hat{g}$ which is a $4\times4$ matrix in Nambu-spin space. This retarded Green function depends on both the normal Green function $\underline{g}$, which is a $2\times2$ matrix in spin space determining spectral properties such as the density of states, and the anomalous Green function $\underline{f}$, which contains information about the superconducting correlations.

To solve this equation, it needs to be complemented by boundary conditions. In our system, we will have both magnetic and spin-orbit coupled interfaces. For the magnetic (also referred to as spin-active) interfaces we have
\begin{align}
    \gLhat \delz \gLhat &= 
    G_0 \com{\gLhat}{\gRhat} 
    + G_1 \com{ \gLhat}{\mhat \gRhat \mhat} 
    \notag\\
    &+ G_{MR} \com{\gLhat}{\anticom{\gRhat}{\mhat}}
    - i G_\varphi \com{ \gLhat}{\mhat}.
    \label{eq:spin_active_bc}
\end{align}
Here, $\hat{m} = \text{diag}(\vecb{m} \cdot \vecb{\tau}, \vecb{m} \cdot \vecb{\tau}^{*})$ where $\vecb{m}$ is a unit vector that describes the interface magnetization, $G_0$ describes the ratio between the barrier resistance and the normal-state resistance of material $L$, $G_{MR}$ describes the magnetoresistance effect of the interface, while $G_1$ originates from the spin-dependent transmission probabilities of a spin-active interfaces. Finally, $G_\phi$ describes the effect of quasiparticles picking up spin-dependent phase shifts as they scatter at the interface. The boundary conditions for $\hat{g}_R$ is obtained by interchanging $L$ and $R$ and multiplying the entire right hand side with $(-1)$. For an in-depth discussion and derivation of these boundary conditions, see \cite{cottet_prb_09, eschrig_njp_15, ouassou_scirep_17}.
We will also need the boundary conditions for a spin-orbit coupled interface \cite{amundsen_prb_19, linder_prb_22}. Defining $\vectau_\parallel = (\rho_x,0,\rho_z)$, these read:
\begin{align}\label{eq:SOC_boundary_conditions}
D\check{g}_R \partial_y \check{g}_R &= T_0^2[\check{g}_L,\check{g}_R] - \tfrac{2}{3} T_1^2p_F^2[\check{g}_R, \vectau_\parallel \check{g}_L\vectau_\parallel] \notag\\
&- mDT_0T_1[\check{g}_R, \{\vectau_{\parallel,x}, \check{g}_L\partial_z \check{g}_L\}] \notag\\
&+ 
Dd\alpha^2[\rho_x, \check{g}_R\rho_x \check{g}_R] + Dd\alpha^2[\rho_z, \check{g}_R\rho_z \check{g}_R]
\end{align}

\noindent where $T_0$ and $T_1$ are phenomenological interface parameters describing respectively spin-independent tunneling and spin-flip tunneling induced by the interfacial Rashba spin-orbit coupling, $m$ is the electron mass, $p_F$ is the Fermi momentum, $\alpha$ quantifies the spin-orbit coupling strength, and $d$ is the thickness of the spin-orbit coupled interface.

\section{Observables}
\label{sc:observables}

Our primary goal is to determine how triplet Cooper pair supercurrents can induce a magnetization in a normal metal by scattering on a spin-orbit coupled interface. To do so, we need the expressions for magnetization and current in quasiclassical theory. In this section, the quasiclassical expressions for the magnetization as well as spin and charge currents are presented. To simplify the analytical study later on we also present the observables expressed in the singlet-triplet decomposition in the weak proximity regime. 

\subsection{Currents}
The quasiclassical expression for current can be found by using the continuity equation  $\partial_t \rho + \nabla \cdot \vecb{j} = S$, where $\vecb{j}$ is the current of a quantity, $S$ is any source-term present, and $\rho$ is the density of the quantity such as charge or spin. In a normal metal, the source term vanishes for the charge current. 

For charge, we have that the density can be written as $\rho = e \braket{\psidag \rhotre \psi }$, and for spin the spin density can be written as  $\braket{\psidag \tfrac{1}{2} \rhotre \vecb{\tau} \psi}$ where $e<0$ is the electron charge and we set $\hbar=1$. 
By using the Heisenberg equation of motion for the creation and annihilation operators $\psi^\dag,\psi$ and writing the expression in terms of the quasiclassical expression we get for the charge current density \cite{belzig_micro_99, eschrig_rpp_15}

\begin{equation}
    \vecb{J} =  \frac{e N_0 D}{4} \int \dd E \Tr \big [\rhotre (\check{g}  \Bar{\nabla} \check{g} )^{K} \big ]. 
\end{equation}

The spin current density is a tensor, with a direction of flow in real space and a polarization direction in spin space, obtained by replacing $\hat{\rho}_3$ with $\frac{1}{2}\hat{\rho}_3 \vecb{\tau}$ 

As mentioned, we will write the expressions in terms of the singlet-triplet decomposition terms $f_s$ and $\vecb{d}$: $\underline{f} = (f_s + \boldsymbol{d}\cdot \boldsymbol{\underline{\vecsigma}})\i\sigma_y$.
By linearizing in $\hat{f}$, assuming $\hat{g} = \rhotre + \hat{f}$, using the relation $\hat{g}^{A} = - \rhotre (\hat{g}^{R})^{\dagger} \rhotre$ and that in equilibrium we have $\hat{g}^{K}  = (\hat{g}^{R} - \hat{g}^{A}) \tanh(\frac{\beta E}{2})$ we get the following expressions for the current density $\vecb{J}$ and spin current density $\vecb{J}_{\! \vecb{s}}$

\begin{align}
    \vecb{J} &= J_0 \int_0^{\infty} \frac{\dd E}{\Delta_0} \xi \tanh(\tfrac{\beta E }{2}) \Re \big ( 
    [  f_s \nabla \tfs -  d_z \nabla \tdz \notag\\
    &-  \dx \nabla \tdx -  \dy \nabla \tdy ] - [\Tilde{\dots}]
    \big ), 
    \label{eq:observables_J}
    \\ \vecb{J}_{ \! s_x} &=  
    J_{ s  0} \int_0^{\infty} \frac{\dd E}{\Delta_0} \xi\tanh(\tfrac{\beta E }{2}) \Im ( 
    [\dy \nabla \tdz - \dz \nabla \tdy ] 
    + [\Tilde{\dots}]
    ),
    \\ \vecb{J}_{ \!s_y} &= 
    J_{ s 0} \int_0^{\infty} \frac{\dd E}{\Delta_0} \xi\tanh(\tfrac{\beta E }{2}) \Im ( 
    [\dz \nabla \tdx - \dx \nabla \tdz ] 
    + [\Tilde{\dots}]
    ),
    \\ \vecb{J}_{\!s_z} &= 
    J_{s0} \int_0^{\infty} \frac{\dd E}{\Delta_0} \xi\tanh(\tfrac{\beta E }{2}) \Im ( 
    [\dx \nabla \tdy - \dy \nabla \tdx ] 
    + [\Tilde{\dots}]
    ),
    \label{eq:observables_Jsz}
\end{align}
where $J_0 =  2 e N_0 D \Delta_0/\xi  $  and $J_{s0} = N_0 D \Delta_0/\xi$. We have written the integral in terms of the dimensionless variable $E/\Delta_0$, where $\Delta_0 = \Delta(T=0)$ is the zero temperature energy gap, and the dimensionless spatial coordinate $z/\xi$. This means that $J/J_0$ are now dimensionless and can be used in the numerical study.

In the following sections, we will also use the charge current divided into which component carries the current. Therefore we introduce the notation
\begin{align}
    \vecb{J}_{ \! f_s} &=  J_0 \int_0^{\infty} \frac{\dd E}{\Delta_0} \tanh(\tfrac{\beta E }{2}) \Re \big ( 
    [  f_s \nabla \tfs]  - [\Tilde{\dots}]
    \big ), 
    \label{eq:observables_Jfs}
    \\
    \vecb{J}_{ \! d_i} &= -   J_0 \int_0^{\infty} \frac{\dd E}{\Delta_0} \tanh(\tfrac{\beta E }{2}) \Re \big ( 
    [   d_i \nabla \tilde{d}_i] - [\Tilde{\dots}]
    \big ), 
    \label{eq:observables_Jdi}
\end{align}
where $\vecb{J}_{\! f_s}$ is the charge current carried by the singlet component and $\vecb{J}_{\! d_i}$ is carried by the $d_i$ component.

\subsection{Magnetization}

The quasiclassical expression for magnetization reads \cite{eschrig_rpp_15}
\begin{equation}
    \vecb{M} = \frac{g \mu_B N_0}{8} \int \dd E \Tr (\hat{\vecb{\tau}} \hat{g}^{K} ).
\end{equation}
where $g$ is the Lande $g$-factor, $N_0$ is the normal-state density of states at the Fermi level, and $\mu_B$ is the Bohr-magneton.
It should be mentioned that this expression does not take into account the contribution from the entire Fermi sea, and is thus not suitable to compute the magnetization of a ferromagnetic metal. It is, on the other hand, suitable for computing any spin-magnetization arising in otherwise non-magnetic materials such as normal metals or conventional superconductors. This holds both for spin accumulations in non-equilibrium systems and proximity-induced equilibrium spin magnetizations.

Once more we want to express the magnetization in terms of the singlet-triplet decomposed components. This can be done similarly to the current, by using the expression for $\hat{g}^{A}$ and the equilibrium expression for $\hat{g}^{K}$.
There are no first-order contributions in the anomalous Green function so we have to take into account the normalization condition, $(\hat{g}^{R})^{2} = 1$. 
By this method, it can be shown that the magnetization, to the second order in the anomalous Green function, reads 
\begin{align}
    M_x &=   M_0 \int_0^{\infty} \frac{\dd E}{\Delta_0} \tanh(\tfrac{\beta E }{2}) \Re(  \tdx \fs -\dx \tfs  ),  
    \label{eq:obsrevables_mx}
    \\M_y &=   M_0 \int_0^{\infty} \frac{\dd E}{\Delta_0} \tanh(\tfrac{\beta E }{2}) \Re(\tdy \fs -  \dy \tfs   ), 
    \label{eq:observables_my}
    \\M_z &=   M_0 \int_0^{\infty} \frac{\dd E}{\Delta_0} \tanh(\tfrac{\beta E }{2}) \Re( \tdz \fs - \dz \tfs   ), 
    \label{eq:observables_mz}
\end{align}
where $M_0 = g\mu_B N_0 \Delta_0$. 

\section{Riccati parametrization}

One particularly convenient way of parametrization the Green function is the Riccati parametrization
\cite{schopohl_prb_95, schopohl_arxiv_98, konstandin_prb_05, jacobsen_prb_15}. 
The Riccati parametrization is advantageous for numerical computation because the parameters are bounded between 0 and 1.
For the purpose of studying systems numerically we will here briefly outline the derivation of the Riccati parametrized Usadel equation as well as giving a detailed derivation of the Riccati parametrized boundary equation in the Appendix, including the effect of a spin-orbit coupled interface, since the latter is not present in existing literature.

The retarded Green function is defined via parameters $N$ and $\gamma$ as follows. 
\begin{equation}
    \hat{g} = 
    \begin{pmatrix}
        N (1 + \gm \tgm)  & 2 N \gm \\
        - \tN \tgm & - \tN (1 + \tgm \gm ) 
    \end{pmatrix}
\end{equation}
$N$ and $\gamma$ will only be used for $2 \times 2$ matrices, so we do not use any special notation to indicate their matrix nature.   
By the normalization condition $\hat{g}^2=1$ it is seen that $N = (1 - \gm \tgm)^{-1}$ and $\tN = (1 - \tgm \gm)^{-1}$. 

A couple of useful identities can be found
\begin{equation}
    N \gm  = \gm \tN, \: \: \tN \tgm = \tgm N.
    \label{eq:riccati_idendtity_one}
\end{equation}

Notice also that
\begin{equation}
    \gm \tgm = 1 - N^{-1}.
\label{eq:riccati_idendtity_two}
\end{equation}

When writing the Usadel equation and the boundary conditions, in particular including the role of spin-orbit coupling, we will have to deal with derivatives. To simplify the notation we therefore introduce $\gm' = \delz \gm$. 
The following way of writing derivatives will also be useful 
\begin{align}
    \delz N = N ( \gm' \tgm + \gm \tgm') N \\
    \delz \tN = \tN ( \tgm' \gm + \tgm \gm' ) \tN \\
    \delz( N \gm)  = N ( \gm' +  \gm \tgm' \gm) \tN \\ 
    \label{eq:riccati_derivative}
    \delz (\tN \tgm) = \tN (  \tgm' + \tgm \gm' \tgm ) N \\
    \label{eq:riccati_derivative_tilde}
\end{align}
all of which can be found by using the identities above. 

The general approach of how to identify the Riccati parameterized Usadel equations and the Kuprianov-Lukichev boundary conditions is thoroughly described in \cite{jacobsen_prb_15}. 
The method consists of writing the terms in Usadel in $4 \times 4$ matrix notation and then taking the upper-right $2 \times 2$ matrix minus the upper-left $2 \times 2$ matrix multiplied by $\gm$. More specifically, one takes $\tfrac{1}{2} N^{-1}([\dots]_{12} - [\dots]_{11} \gamma )$ of the matrix equation $[\dots]$, where the subscript indicates a block matrix. 
Doing so, one manages to eliminate terms such that $\delz^{2} \gm$ or $\delz \gm$ terms can be separated out. 

The final result for the Riccati parameterized Usadel equation for a ferromagnet reads
\begin{equation}
    \delz^{2} \gm = - 2 i E \gm - i \vecb{h} \cdot (\tau \gm - \gm \tau^{*}) - 2 \gm' \tN \tgm \gm',
\end{equation}
where for a normal metal, the only adjustment needed is to set $\vecb{h} =0$. It can be shown that the Riccati parameterized bulk BCS singlet superconductor solution is
\begin{equation}
    \gamma_{\text{BCS}} = 
    \begin{pmatrix}
        0 & b\e{i\phi} \\
        -b\e{i\phi} & 0
    \end{pmatrix}, 
\end{equation}
where $\phi$ is the phase and
\begin{equation}
    b = 
    \begin{cases}
        \frac{\Delta}{E + i \sqrt{\Delta^{2} - E^{2}}}, \: \: \text{for} |E| < \Delta \\
        \frac{\Delta \sgn(E)}{ |E| + \sqrt{E^2 - \Delta^2} } \: \: \:\text{for} |E|> \Delta.
    \end{cases}
\end{equation}

\subsection{Riccati parametrization of boundary conditions}

When deriving the Riccati parametrizing of the specific boundary conditions used in this work, it is useful to note that many of the terms both in the spin-orbit coupling and spin-active boundary conditions have the same form. 
Here a practical method for Riccati parametrizing this type of terms is presented in order to simplify the calculations. 

The form, which many of the terms take, is
\begin{equation}
     [\ghat_L , \Uhat],
    \label{eq:general_bc_term}
\end{equation}
where $\Uhat$ is a matrix whose exact form depends on the specific boundary conditions. In general, we write $\Uhat$ as 
\begin{equation}
    \Uhat =
    \begin{pmatrix}
        \Ubar_{11} & \Ubar_{12} \\
        \Ubar_{21} & \Ubar_{22}
    \end{pmatrix}.
    \label{eq:Uhat}
\end{equation}

By Riccati parametrization, it can be found from the left-hand side of the boundary equation $\hat{g}\partial_z \hat{g}$ that
\begin{equation}
    \frac{1}{2} N_L^{-1} \big ( [\ghat_L \delz \ghat_L]_{12} - [\ghat_L \delz \ghat_L]_{11} \gamma_L \big )  = \delz \gamma_L,
    \label{eq:left_hand_side_bc_riccati}
\end{equation}
as seen in \cite{jacobsen_prb_15}. The subscript tells which block of the matrix that is meant, as in Equation \eqref{eq:Uhat}. 

To obtain the complete boundary conditions, we also have to perform the same operation on the right-hand side as on the left side. This means we need to take $\tfrac{1}{2} N^{-1}_L ( [\dots]_{12} - [\dots]_{11} \gamma_L )$ of every term on the right-hand side of the boundary conditions. Thus we start by finding a procedure for all terms that comes in the form of Eq. \eqref{eq:general_bc_term}. 

\begin{equation}
\begin{aligned}
    & \frac{1}{2} N_L^{-1} \big ( [\ghat_L , \Uhat]^{1,2} - [\ghat_L , \Uhat]^{1,1} \gamma_L \big ) \\
    &= \frac{1}{2} N_L^{-1} \big (  \gbar_L \Ubar_{12} + \fbar_L \Ubar_{22} - \Ubar_{11} \fbar_L  + \Ubar_{12} \tgbar_L  \notag\\
    &- ( \gbar_L \Ubar_{11} + \fbar_L \Ubar_{21} - \Ubar_{11} \gbar_L  + \Ubar_{12} \tfbar_L )\gamma_L    \big ).  \\
\end{aligned}
\end{equation}

As a next step we collect the terms with the same $U_{ij}$ matrices and insert $\fbar = 2N\gamma$, $\gbar = 2N - 1$.  The $\Ubar_{11}$ terms can be written as 

\begin{equation}
\begin{aligned}
    &\;\tfrac{1}{2} N_L^{-1} ( - \Ubar_{11} \fbar_L - \gbar_L \Ubar_{11} \gamma_L + \Ubar_{11} \gbar_L \gamma_L ) \\ 
    &= \tfrac{1}{2} N_L^{-1} (- \Ubar_{11} 2 N_L \gamma_L - (2N_L -1) \Ubar_{11} \gamma_L +  \Ubar_{11} (2N_L -1 ) \gamma_L ) \\ 
    &= - \Ubar_{11} \gamma_L. 
\end{aligned}
\end{equation}

In the same manner the $\Ubar_{12}$ terms becomes
\begin{equation}
\begin{aligned}
    &\;\tfrac{1}{2} N_L^{-1} \big ( \gbar_L \Ubar_{12} + \Ubar_{12} \tgbar_L - \Ubar_{12} \tfbar_L \gamma_L \big )\\
    &= \tfrac{1}{2} N_L^{-1} \big (  (2N_L -1) \Ubar_{12} + \Ubar_{12} (2\tN_L- 1) - \Ubar_{12} 2 \tN_L \tgamma_L \gamma_L \big ) \\
    &= \Ubar_{12}.
\end{aligned} 
\end{equation}

The $\Ubar_{21}$ term should also be written in terms of $\gamma$ as 
\begin{equation}
\begin{aligned}
    & \tfrac{1}{2} N_L^{-1} ( - \fbar_L \Ubar_{21} \gamma_L ) \\
    & = - \gamma_L \Ubar_{21} \gamma_L. \\
\end{aligned}
\end{equation}

Finally, the $\Ubar_{22}$ term can be written as
\begin{equation}
\begin{aligned}
    & \tfrac{1}{2} N_L^{-1} (\fbar_L \Ubar) \\ 
    &= \gamma_L \Ubar_{22}. 
\end{aligned}
\end{equation}

Putting everything together we get
\begin{equation}
\begin{aligned}
     \frac{1}{2} &N_L^{-1} \big ( [\ghat_L , \Uhat]^{1,2} - [\ghat_L , \Uhat]^{1,1} \gamma_L \big )  = - \Ubar_{11} \gamma_L +  \Ubar_{12}  \notag\\
     &- \gamma_L \Ubar_{21} \gamma_L + \gamma_L \Ubar_{22}.
     \label{eq:general_bc_term_riccati}
\end{aligned}
\end{equation}

We note that this also can be used for the Kuprianov-Lukichev boundary conditions \cite{kupriyanov_zetf_88}, where one would have $\Uhat = G_0 \ghat_{R}$. From this, and using the identities in Eqs. \eqref{eq:riccati_idendtity_one} and \eqref{eq:riccati_idendtity_two},
the Kuprianov-Lukichev boundary conditions can be found to be
\begin{align}
    \delz \gm_L = G_0 (1 - \gm_L \tgm_R ) N_R ( \gm_R - \gm_L )  \\
    \delz \gm_R = G_0 (1 - \gm_R \tgm_L ) N_L ( \gm_R - \gm_L ).  
\end{align}
In the Appendix, both the spin-orbit coupling and spin-active boundary conditions will be written in the Riccati parameterized form using the method described here.

\section{Setup}
\label{sc:setup}

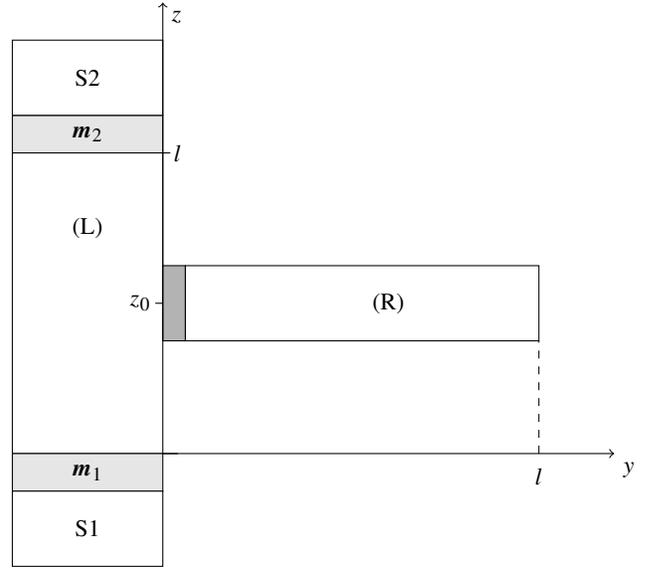
\begin{figure}
    \centering
    \begin{tikzpicture}
        \filldraw[fill = black!10, draw = black] (-2,2) rectangle (0,2.5);
        \filldraw[fill = white, draw = black] (-2,-2) rectangle (0,2);
        \filldraw[fill = black!30, draw = black] (0,-0.5) rectangle (0.3,0.5);
        \filldraw[fill = white, draw = black] (0.3,-0.5) rectangle (5,0.5);
        \filldraw[fill = black!10, draw = black] (-2,-2.5) rectangle (0,-2);
        \filldraw[fill = white, draw = black] (-2,2.5) rectangle (0,3.5);
        \filldraw[fill = white, draw = black] (-2,-2.5) rectangle (0,-3.5);
        \node at (-1,1) {(L)};
        \node at (3,0) {(R)};
        \draw[->] (0,-2) -- (6,-2) node[anchor=north west] {$y$};
        \draw[->] (0,-0.5) -- (0,4) node[anchor=north west] {$z$};
        \draw (0,-2) -- (0.2,-2);
        \node at (5,-2.3) {$l$};
        \node at (-1,3) {S2};
        \node at (-1,-3) {S1};
        \node at (-1,2.25)  {$\vecb{m}_2$};
        \node at (-1,-2.25) {$\vecb{m}_1$};
        \node at (-0.3,0) {$z_0$};
        \node at (0.2,2) {$l$};
        \draw[dashed] (5,-2) -- (5,-0.5);
        \draw (-0.1,0) -- (0,0);
        \draw (0,2) -- (0.1,2);
    \end{tikzpicture}
    \caption{(Color online) The system in which supercurrents and induced magnetization is investigated. The material to the left, (L), is the material in which the supercurrent will flow. This material will be either a ferromagnet or a normal metal. To get the current flowing, two conventional superconductors are connected to (L), and a phase difference between them is applied. 
    In between the superconductors and (L), spin active interfaces are included for the purpose of creating triplet Cooper pairs and thereby also triplet supercurrents. A normal metal, (R), is connected to (L) through a spin-orbit coupled interface. This material borders to vacuum at $y = l$.}
    \label{fig:system}
\end{figure}

\begin{figure}
    \centering
    \begin{tikzpicture}
    \coordinate (origo) at (0,0);
    \coordinate (a) at (1,0);
    \coordinate (b) at (2,3);
    \coordinate (c) at (3,2);
        \draw[->] (0,0) -- (4,0) node[anchor = south west]{$x$};
        \draw[->] (0,0) -- (0,4) node[anchor = north east]{$y$};
        \draw[->] (0,0) -- (2,3) node[anchor = north west]{$\vecb{m}_2$};
        \draw[->] (0,0) -- (3,2) node[anchor = north west]{$\vecb{m}_1$};
        \pic [draw,->, "$\alpha$",black,thick,angle radius=1cm] {angle = a--origo--c};
        \pic [draw,->,"$\theta$",black,thick,angle radius=1.5cm] {angle = c--origo--b};
    \end{tikzpicture}
    \caption{(Color online) The figure shows the definition of the angles $\alpha$ and $\theta$. $\alpha$ is defined as the angle between the $x$-axis and the interface magnetization of in the first interface $\vecb{m}_1$. 
    $\theta$ is the angle between the two interface magnetizations. 
    }
    \label{fig:interface_magnetization_angles}
\end{figure}
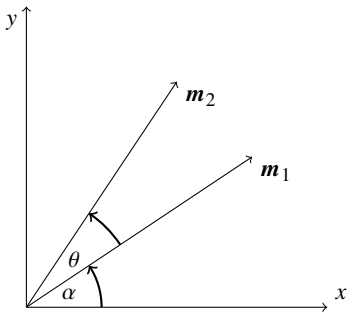
The system we are studying is illustrated in Fig. \ref{fig:system}. 
The region in which the supercurrents will flow is drawn to the left and will therefore be referred to as (L).
Similarly, the region where induced magnetization will be studied is drawn to the right and is called (R). 
In this study, we let the region (R) be a normal metal, and (L) will either be a normal metal or a ferromagnet depending on the situation. The $z$ position where (R) is connected to (L) is called $z_0$. In most situations, we use $z_0 = l/2$, and it will be specified when other values are used. The (L) and (R) regions are chosen to have the same length, but this is not required for the effects predicted here.
The grey region between (L) and (R) is the spin-orbit coupled material, for which the spin-orbit coupling boundary conditions will be used. 
At $z = 0$ and $z = l$, conventional BCS superconductors, S1 and S2, are attached to the (L) material. These superconductors are the sources for Cooper pairs in the rest of the system. Between (L) and the superconductor, spin-active interfaces are introduced and are marked as grey regions in the figure. The interface magnetizations of the spin-active interfaces are described by unit vectors $\vecb{m}_1$ and $\vecb{m}_2$. 

To create a pure singlet charge current, the interface magnetizations are switched off ($\vecb{m}_1 = \vecb{m}_2 = 0$) which is equivalent to using the regular Kuprianov-Lukichev boundary conditions. Furthermore, (L) is a normal metal in the singlet current case. As the superconductors S1 and S2 only contain singlet Cooper pairs, no triplets will be induced in (L) in this scenario. 
We want a supercurrent to flow through the (L) region. This is achieved by applying a phase difference, $\phi$, between S1 and S2. 

To create the triplet charge current and the spin current, the interface magnetizations are switched on. The triplet charge current is created in the same way as in the singlet case: by applying a phase difference.

\begin{figure}[t!]
    \centering
    \includegraphics[width = 0.4 \textwidth]{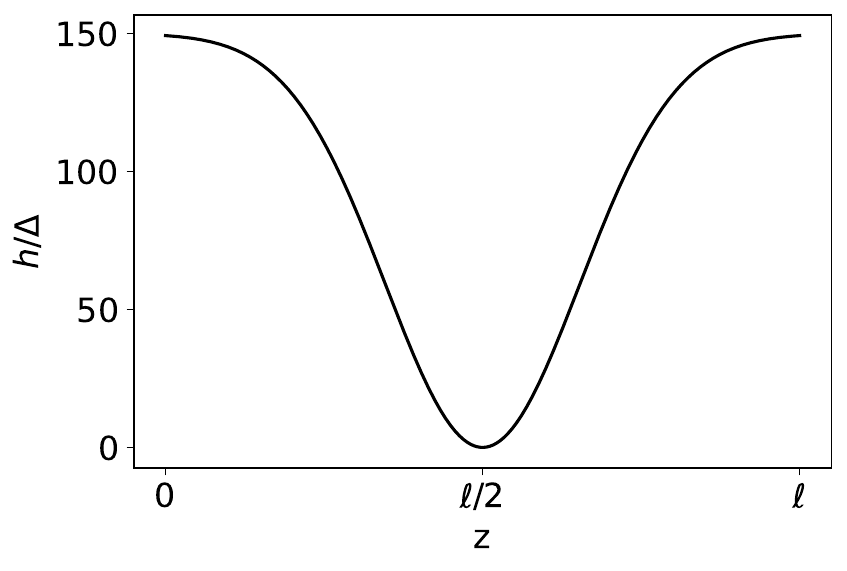}
    \caption{(Color online) The spatial profile of the exchange field function $h(z)$ which is applied to the left material (L) in the triplet current cases. This ensures that only the long-ranged triplet component survives at the contact with the normal metal through the Rashba spin-orbit coupled interface. In an experiment, one would use a strong ferromagnet/normal metal bilayer to achieve an exchange field spatial profile serving the same purpose. In this work, the spatial profile of $h$ is continuous rather than abrupt for numerical convenience, as one can then evaluate the Usadel equation in a single layer.} 
    \label{fig:exchange_field}
\end{figure}

For the purpose of discovering effects caused solely by triplets, a ferromagnetic exchange field is included in the material (L). Because of the exchange field, the singlet becomes short-ranged and dies out rapidly in the (L) region. 
In an experimental setup, it would be of importance to separate the intrinsic magnetization coming from an exchange field and the magnetization induced by supercurrents. 
Therefore the exchange field is modeled to be spatially varying in (L) such that it is zero in the middle region, but large at the sides as shown in Fig. \ref{fig:exchange_field}. 
In practice, this can be realized by attaching thin ferromagnetic regions with a strong exchange field right next to the superconductors and then having a long normal metal region separating the ferromagnets. In this way, the singlets and short-range triplets are filtered out by the thin, strongly polarized ferromagnetic regions, whereas the long-range triplets produced in the ferromagnetic region remain and can propagate through the normal metal.
More specifically, it is the triplet component that is spin-neutral in the exchange field orientation $\vecb{d} || \vecb{h}$ that is short-ranged, and the others are long-ranged.

As we will discuss in the following analytical study, the difference between the $d_z$ and $d_y$ component is quite insignificant relative the spin-orbit coupled interface, and it is instead the $d_x$ component that is the most relevant. 
Therefore, we focus most of the discussion on the case where the interface magnetizations lie in the $xy$-plane, and the exchange field points in the $z$-direction, $\vecb{h} = (0,0,h(z))$. 
We do, however, include rotation of both $\vecb{m}_1$ and $\vecb{m}_2$ with an angle $\alpha$ around the $z$-axis, and the angle between $\vecb{m}_1$ and $\vecb{m}_2$ which we call $\theta$.
These angles are illustrated in Fig. \ref{fig:interface_magnetization_angles}. We also note here that the directions chosen are advantageous for an experimental setup, as rotating the interface magnetizations in-plane is a simpler task than driving them out of the $xy$-plane. 

We also remark that rotating both interface magnetizations, $\vecb{m}_1$ and $\vecb{m}_2$, by an angle $\alpha$ is equivalent to rotating material (L) around the $z$-axis. Thus, rotating the interface magnetization or attaching the (R) region to the (R) region at different angles corresponds to the same physical system.

In the numerical study, we have used the interface parameters $\prefnull =   0.2, \prefen = \prefto = \preftre = 0.1$, $G_0 = 1/\xi_s$.
To create pure charge currents, we set $P=0$ and $G_\phi = 3G_0$. We also checked that the results are qualitatively similar if one instead uses $P \neq 0, G_\phi=0$. If both $P$ and $G_\phi$ are finite, spin supercurrents in addition to charge supercurrents flow in the system.
Furthermore, a combination of the length of material (L), $l$, that is short enough to preserve some of the triplets, and strength of $\vecb{h}$ that is strong enough to remove the singlet had to be found. 
Setting the length of material $l$ to eight times the bulk superconducting coherence length $l= 8 \xi_s  = 8 \sqrt{D/\Delta_0}$, and $h(z)$ as in Figure \ref{fig:exchange_field} was found to work well.

Finally, we emphasize that our main aim here is to determine the qualitative behavior of the spin-signal induced by the charge and spin currents, such as when it exists and how it changes depending on the polarization of the triplet Cooper pairs, and not predict its precise magnitude, which will depend on the material choices.

\subsection{Numerical method}
In general, the materials on both sides of an interface are affected by each other. However, here it is assumed for simplicity that the inverse proximity effect that material (L) induces in S1 and S2 is negligible, and only the effect from the superconductor on the material (L) is considered. The superconductors are thus assumed to be the bulk superconductors. Such an approximation is valid when one of the materials is much more disordered than the other \cite{bergeret_rmp_05}. The bulk superconductor Green function can by this assumption be used directly in the boundary conditions used to solve system (L). 
Similarly, the effect from (R) on (L) is neglected, and the solution for system (L) at $z = z_0$ is used directly into the spin-orbit coupling boundary conditions for (R). 
Furthermore, the materials are individually modeled as one dimensional, meaning that material (L) is extended in the $z$-direction and (R) in the $y$-direction. 

The system which has to be solved is now the one-dimensional Usadel equation in two regions.    
This is a second-order differential equation for two variables $\gamma$ and  $\tgm$, and to solve it numerically we rewrite it as four first-order differential equations writing

\begin{equation}
    \begin{pmatrix}
        \gamma' \\
        \gamma'' \\
        \tgm' \\ 
        \tgm''
    \end{pmatrix}
    = f 
    \begin{pmatrix}
        \gamma \\
        \gamma' \\
        \tgm \\
        \tgm' 
    \end{pmatrix},
\end{equation}
where $f(\ldots)$ is a function that returns the derivative of the input. The function will thus return $\gamma'$ as the derivative for $\gamma$ and use the Riccati parameterized Usadel equation to find the derivative of $\gamma'$ and similarly for $\tgm$. 
Thus we have a system of $16$ complex connected differential equations, four elements in each of the matrices $\gm, \gm', \tgm $, and $\tgm'$. 
The boundary conditions give restrictions to $\gamma'$ and $\tgm'$ on each side of the material. 

To solve the system we have used the boundary value problem solver from SciPy \cite{noauthor_scipyintegratesolve_bvp_nodate}. To stabilize the solver the real and imaginary part are split such that the 16 complex equations become 32 real ones. 
To increase numerical stability,  inelastic scattering is also included by adding a small imaginary component to the energy, which here is set to $\delta/\Delta_0 = 0.01$ as used in \cite{ouassou_prb_17}. 
This imaginary component is  referred to as the Dynes parameter and is often used to model experiments \cite{dynes_prl_78}.
In essence, it has the effect of broadening the spectral features, such as the peaks of the Green functions that occur at $E =0$ and $E = \Delta$. 
As mentioned, the interface magnetizations are rotated in the $xy$-plane in the triplet cases.
A small trick was made for solving the system with these rotations. 
To save computation time, the Usadel equation only has to be solved once in material (L) for one given $\theta$ and one given set of interface parameters. The angle $\alpha$ can simply be taken into account by rotating the triplet components $\dxL$ and $\dyL$, such that 
\begin{equation}
\begin{aligned}
    \dxL(\alpha)  &= \dxL(0) \cos(\alpha) + \dyL(0) \sin(\alpha), \\
    \dyL(\alpha)  &= -\dxL(0) \sin(\alpha) + \dyL(0) \cos(\alpha).
\end{aligned}
\end{equation}
This means that instead of solving the system in (L) for every $\alpha$, we solve it once and then rotate the solution to proceed studying the material (R). 
For the material (R), however, the Usadel equation needs to be solved separately for every value of $\alpha$. We have verified the integrity of our numerical method by \eg checking that current conservation is satisfied and that we reproduce known results from the literature, such as the induced magnetization from a singlet charge current in \cite{linder_prb_22}.

\section{A brief analytical study}
\label{sc:analytical_study}

To gain more physical insight before proceeding to the numerical results, we here analyze the weak proximity effect regime where the equations can be linearized in the anomalous Green function. 
We will see that there is a clear relation between singlet charge current, $J_{f_s}$, and the induced magnetization in the $x$-direction, $m_x$. 
A similar relation also exists between a triplet charge current carried by the $\dx$ triplet, $J_{d_x}$, and $m_x$. 

In the weak proximity regime, we assume that the quasiclassical Green function is close to the normal metal solution, $\ghat_{N} = \rhotre$, but with a small superconducting part, $\fhat$, induced by the proximity superconductors. We thereby assume that we can use the weak proximity solution $\hat{g} \approx \rhotre + \hat{f}$. Using the singlet-triplet decomposition, the Rashba SOC boundary conditions can be written as follows,
keeping only the terms of the first order in $\fhat$ 
\begin{align}
    \dely \fsR =&  -\prefto \delzdxL \notag\\
    &- 2 (\prefnull - 2 \prefen) (\fsL - \fsR)
    \label{eq:Rasha_BC_fs}
    \\
    \dely \dzR =& - (8 \preftre - 2 \prefnull + 4 \prefen) \dzR \notag\\
    &- (2 \dzL) (\prefnull + 2 \prefen)
    \label{eq:Rashba_BC_dz}
    \\
    \dely \dxR =&- 4 \delzfsL \prefto - (4 \prefen + 4 \preftre - 2 \prefnull) \dxR \notag\\
    &-  (2 \dxL) \prefnull
    \label{eq:Rashba_BC_dx}
    \\
    \dely \dyR =&- (4 \prefen + 4 \preftre - 2 \prefnull) \dyR - 2 \prefnull \dyL 
    \label{eq:Rashba_BC_dy}
\end{align}

From this, we see that there is a link between $\fsR$ and $\delzdxL$ and the other way around between $\dxR$ and $\delzfsL$. 
On the other hand, to first order, $\dyR$ can only be induced by $\dyL$ and $\dzR$ only by $\dzL$.

Furthermore, we can use the solution to the linearized Usadel equation in a normal metal: 
\begin{equation}
\begin{aligned}
    f_s = A_s \ex{- k y} + B_s \ex{ k y}, \\ 
    d_x = A_x \ex{- k y} + B_x \ex{ k y},
    \label{eq:linearized_normal_metal_solution}
\end{aligned}
\end{equation}
where $k = \sqrt{-2iE/D}$ and $A_s,A_x,B_s$, and $B_s$ are constants that have to be determined using the boundary conditions.
If we assume that $l \to \infty$, the $B$ factors have to be zero in order to avoid a diverging function.

\subsection{Singlet current}
We start by discussing the singlet charge current case. 
In this scenario, there are no triplet components present in (L). 
Thus, there is no induced $\dy$ or $\dz$ on the right side of the interface. 
Furthermore, there is no $\dx$ and thus no $\dely \dx$ on the left side. Assuming for simplicity $l \to \infty$, from Eq. (\ref{eq:Rasha_BC_fs}) and Eq. (\ref{eq:Rashba_BC_dx}) we then see that  
\begin{align}
& \fsR \propto \fsL, 
\\& \dxR \propto \delz \fsL.   
\end{align}
since in this case $\partial_y f_{s,R} \propto f_{s,R}$ and $\partial_y d_{x,R} \propto d_{x,R}$.
Note that the tilde-conjugated components have the same proportionality between the left and right sides.
Thus, it follows that 
\begin{equation}
    \fsL \delz \tfsL - \tfsL \delz \fs \propto  \fsR \tdxR - \tfsR \dxR. 
\end{equation}
From 
Eq. \eqref{eq:observables_Jfs} and Eq. \eqref{eq:obsrevables_mx}
we see that the real part of the left side in the expression above is exactly what occurs in the expression for a singlet current in material (L), $J_{f_s}$. We also see that the real part of the right side in the expression is exactly what occurs in the induced magnetization expression $m_{x,R}$.

From the analytical expression, we conclude that there is a clear connection between induced magnetization and the singlet charge current. 
We note, however, that $\fsL \delz \tfsL - \tfsL \delz \fsL$ and $\fsR \tdxR - \tfsR \dxR$ might not have the same phase, such that the real part of these expressions might not be directly proportional.

\subsection{Triplet charge current}
Next, considering a triplet current we see that the current $J_{d_y}$ and $J_{d_z}$, which contains varying $d_y$ and $d_z$ components in (L), do not induce a singlet in (R), at least to the first order in $f$.
We can thus conclude that $\dy$ and $\dz$ induce no magnetization in (R). 
For the last type of triplet charge current, $J_{d,x}$ the same argumentation as for the singlet charge current applies. The difference is that in this case we have

\begin{align}
    &\dxR \propto \dxL,
    \\&\fsR \propto \delz \dxL.
\end{align}
This gives 
\begin{equation}
    \dxL \delz \tdxL - \tdxL \delz \dx \propto  \fsR \tdxR - \tfsR \dxR. 
\end{equation}
Seemingly, $J_{d_x}$ and $J_{f_s}$ then induce the same magnetization. 

A triplet charge current does, however, not have to be carried by a pure $d_x$, $d_y$, or $d_z$. In general, we can have an arbitrary $\vecb{d}$-vector carrying a pure charge current.
Keeping to the situation where the triplet vector $\vecb{d}$ points in the $xy$-plane, we would get a current carried partially by $d_x$ and partially by $d_y$. 
In this case, the part carried by $d_x$ induces a singlet component in (R), whereas the $d_y$ part induces a $d_y$ in (R) as well. 
By this argument, a magnetization in the $y$-direction would also be induced. 
Thus we see that with a triplet charge current, magnetization can also be induced in more than one direction. This is contrary to the singlet charge current. 

\subsection{Spin current}
We also briefly comment on how the induced magnetization changes when there is not only a charge supercurrent, but also a spin supercurrent flowing in (L). From section \ref{sc:observables} we see that there is no $\dx$ component involved in the $x$-polarized current $J_{s_x}$. From this spin current, no singlet can be induced in (R), and thus also no magnetization. Both for the $y$-polarized spin current, $J_{s_y}$, and the $z$-polarized spin current $J_{s_z}$, a $\dx$ component is involved. Since the $\dy$ and $\dz$ components are treated similarly by the spin-obit coupled interface, we settle for only studying $J_{s_z}$. Presumably, $J_{s_y}$ would induce a rotated, but similar magnetization.

The $z$-polarized spin current, $J_{s_z}$ is seen in 
Eq. \eqref{eq:observables_Jsz}
to be dependent on the imaginary part of the following expression
\begin{equation}
      \dxL \delz \tdyL - \dyL \delz \tdxL + \tdxL \delz \dyL - \tdyL \delz \dxL.  
\end{equation}
Notice that from the linearized spin-orbit coupled boundary conditions, it is found that 
\begin{equation}
    \dyL \delz \tdxL  \propto \dyR \tfsR. 
\end{equation}
Notice also that the real part of the right side also occurs in the $m_y$ expression. 
The relation here is much less direct than in the charge current and $m_x$ case. In the expression for $m_y$ the term and the tilde-conjugated of the term are subtracted from each other, but in the $J_{s,z}$ expression they are added together. 
Furthermore, $m_y$ depends on the real part of the expression whereas $J_{s_z}$ depends on the imaginary part.
It therefore follows that no clear connection exists between a pure spin current and an induced magnetization via the spin-orbit coupled interface.

\subsection{Linearized spin-active boundary conditions}
Before proceeding to the numerical study, we also look at the linearized spin-active boundary conditions in order to understand how the currents are created. 
These boundary conditions will only be used in interfaces where one side is a singlet superconductor. We consider below for concreteness the right interface and thus remove all triplets on the right side of the interface of the equations below. 
The linearized equations in the singlet-triplet decomposition notation read

\begin{figure}[t!]
    \centering
    \includegraphics[width = 0.45\textwidth]{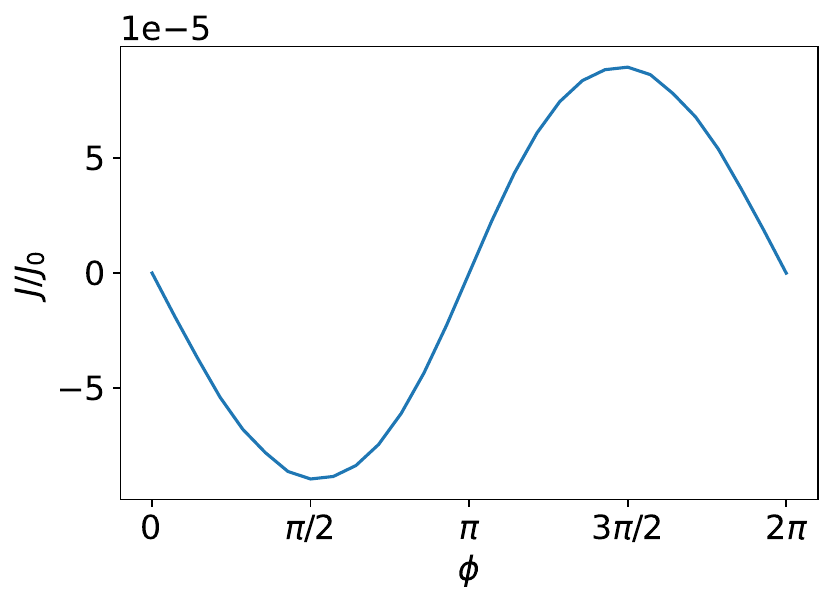}
        \caption{(Color online) The supercurrent induced in material (L) as a function of the phase difference between S1 and S2, $\phi$. }
    \label{fig:singlet_current_current_and_mangetization}
\end{figure}

\begin{figure}[t!]
    \centering
    \includegraphics[width = 0.45\textwidth]{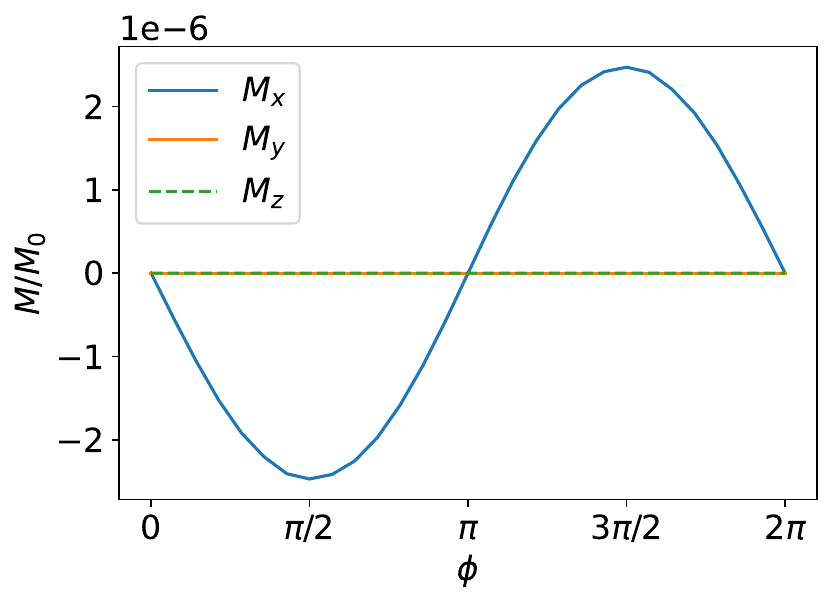}
    \caption{ (Color online) The induced magnetization in (R), right next to the interface SOC interface at $y = 0$, in the presence of a supercurrent flow in region (L) due to a phase difference $\phi$}.
    \label{subfig:singlet_Mx_My_Mz}
\end{figure}

\begin{align}
    \delz \fsL =& -2 \vecb{m}^2 (\fsL + \fsR) G_1 + 2 G_0 \fsR - 2 G_0 \fsL 
    \notag\\
    &- 2 G_{\phi} (\dxL m_x + \dyL m_y + \dzL m_z). \\
    \delz \dxL =& (-\vecb{m}^2 G_1 - 2 G_0)\dxL + (4i\dyL m_z - 4i\dz m_y) G_{\text{MR}} \notag\\
    &- (2 G_{\phi}) \fsL m_x,
    \\
    \delz \dyL =& (-2\vecb{m}^2 G_1 - 2 G_0)\dyL + ( 4i\dz m_x - 4 i m_z\dx )G_{\text{MR}} \notag\\
    &- (2 G_\phi)\fsL m_y,
    \\
    \delz \dzL =& (-2 \vecb{m}^2 G_1 - 2 G_0)\dzL + (4 i \dx m_y - 4 i\dy m_x) G_{\text{MR}} \notag\\
    &- (2 G_\phi)\fsL m_z.
\end{align}

For studying the triplet scenarios, material (L) will be a ferromagnet. Except for when we want to determine the effect of a current $J_{d_z}$,  
the exchange field will be oriented in the $z$-direction which induces a $d_z$ component in (L). As mentioned, this $d_z$ triplet will then be short-ranged, and thus we focus this discussion on the $d_x$ and $d_y$ components induced by the spin-active boundary.

The terms that create the $\dx$ and $\dy$ triplets are the $G_\varphi$- and $G_{\text{MR}}$-terms. If we only have the $G_\phi$ term, we see that the induced $\vecb{d}$ is parallel to the interface magnetization. If we however, turn off $G_\phi$ but turn on $G_{\text{MR}}$ (by letting $P \neq 0$) we see that $\vecb{d}$ is orthogonal to the interface magnetization, since $d_z$ will already be present because of the exchange field in (L). 

If we thus let the interface magnetization be parallel to each other and only include one of the terms such that we have either $G_\phi = 0$ or $P = 0$ the behavior of the current closely resembles that of a conventional Josephson junction, 
except that it is only the triplet and not the singlet that is long-ranged. 
Applying a phase difference between the BCS superconductors S1 and S2, it is reasonable to expect that in this case we can create a pure triplet charge current in (L).
As we, in that case, are able to create only one long-ranged triplet component we know there can not be any spin current. When both $G_\phi$ and $G_{\text{MR}}$ are present, or $\theta \neq 0$, we see that both $d_x$ and $d_y$ will be created, and we can in general also find a spin supercurrent.

\section{Numerical results}

\subsection{Singlet charge current}

The proximity effect and magnetization induced by the singlet supercurrent is explored by removing the interface magnetizations and the exchange field in (L). We numerically determine the magnetization in the full proximity effect regime, using the non-linear Riccati-parametrized equations, and compare it to the results expected from the analytical treatment. Further, the role of the symmetry of the anomalous Green function under the $\tilde{\ldots}$ operation is discussed. Finally, we discuss the spatial dependence of the magnetization induced in the normal metal (R) and also check the temperature dependency.

\begin{figure}
    \centering
    \includegraphics[width = 0.5 \textwidth]{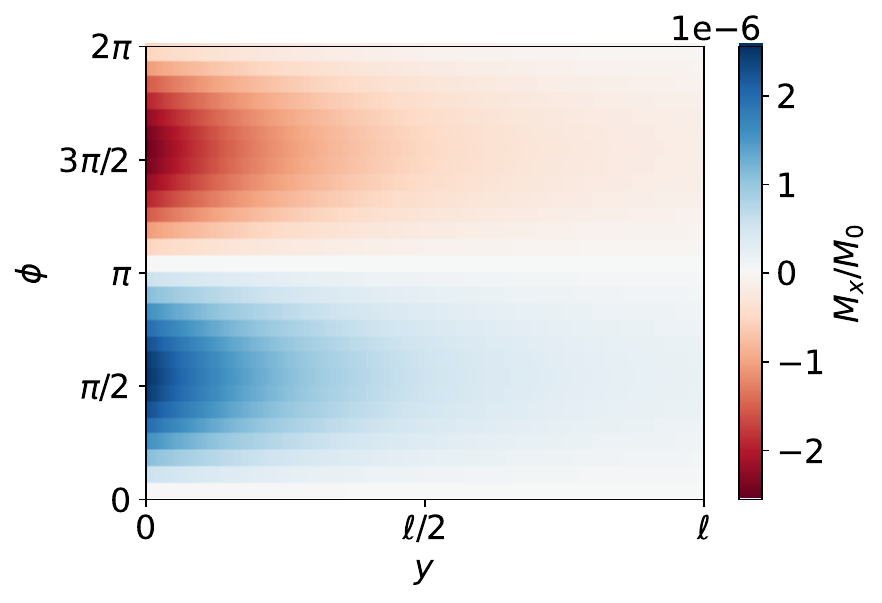}
    \caption{(Color online) The supercurrent-induced magnetization in (R) as a function of phase difference $\phi$ and position $y$ in the zero temperature case. }
    \label{fig:singlet_mx_at_Tzero}
\end{figure}

The current and magnetization as a function of $\phi$ are shown in Figs. \ref{fig:singlet_current_current_and_mangetization} and \ref{subfig:singlet_Mx_My_Mz}. The magnetization shown in the figure is evaluated at $y=0$.
As expected from the analytical study, no magnetization was induced in the $y$ or $z$ direction. 
Furthermore, it is seen from the figure that the induced $m_x$ in (R) is proportional to the singlet current in (L). 
We remark that this is consistent with Ref. \cite{linder_prb_22}, who found from using an effective model for the Green function in (L), that $m_x \propto J$. 
The effective model used in the paper has a disappearing derivative of $f_s$ when $J=0$ as it does not consider that the absolute value of $f_s$ can be decaying. 
Inside a normal metal, however, we expect the Cooper pair wavefunction to decay away from the superconductor interfaces. 
Even when the current is zero, the singlet component can therefore still have a finite derivative. Therefore, it is natural to ask whether the disappearance of the magnetization at $J = 0$ is caused by the spatial gradient of the $f_{s,L}$ and $d_{x,L}$ components vanishing at $\phi = 0, \pi$ or caused by the symmetry properties under the $\tilde{\ldots}$ operation of the triplet and singlet component. We have performed this analysis in the Appendix, and the conclusion is the $\tilde{\ldots}$-operation symmetry, which is influenced by whether or not a supercurrent flows, causes the magnetization to vanish at $\phi = 0, \pi$.
Note that the absence of induced magnetization does not necessarily imply the absence of a triplet being induced. The supercurrent-induced magnetization decays montonically in the (R) normal metal, as shown in Fig. \ref{fig:singlet_mx_at_Tzero}.

Finally, we consider the temperature dependence of the supercurrent-induced magnetization. The quasiclassical magnetization from Eq. \eqref{eq:obsrevables_mx} has a factor $\tanh(\beta E/2)$ in the integrand. 
Varying the temperature changes how the contribution
\begin{equation}
    \Re(\fs(E)\tdx(E) - \tfs(E)\dx(E))
\end{equation}
is weighted with respect to energy. A relevant question to ask is therefore how the temperature affects the magnetization. The temperature-dependence of the energy gap must be taken into account.
We chose a standard interpolation formula which is valid for $T \in (0,T_c)$ \cite{senapati_natmat_11}:

\begin{equation}
    \Delta (T) = \Delta(0) \tanh \left (1.74 \sqrt{\tfrac{T_c}{T} - 1} \right ),
\label{eq:gap_interploation}
\end{equation}

where $T_c$ is the critical temperature.  
We also use the BCS relation between the zero temperature gap and the critical temperature, $\frac{\Delta_0}{T_c} = 1.76$.  

\begin{figure}
    \centering
    \includegraphics[width = 0.5\textwidth]{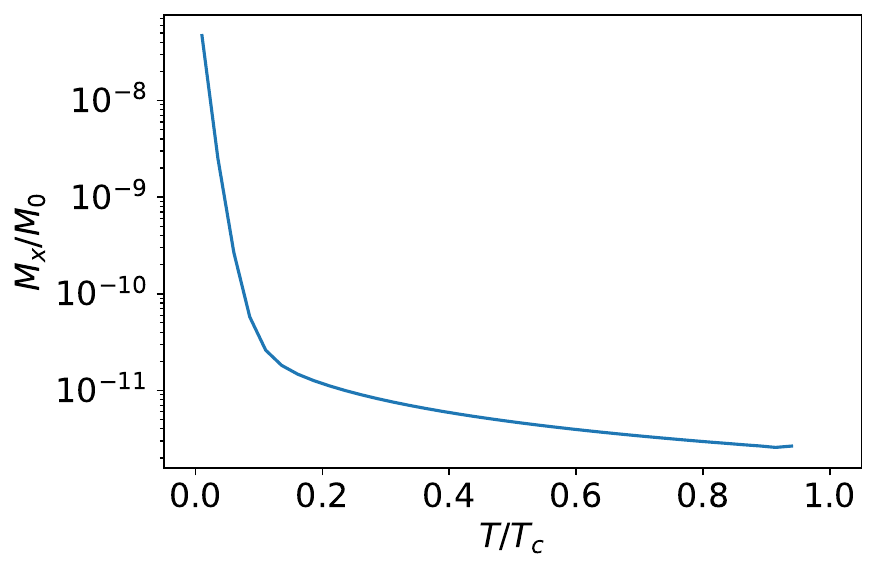}
    \caption{(Color online) The magnetization  evaluated at $y =l$ as a function of temperature. We used $\phi=\pi/4$.}
    \label{fig:mx_of_T}
\end{figure}

Fig. \ref{fig:mx_of_T} shows the magnetization evaluated at $y = l$ as  a function of temperature. Although the integrand in general oscillates as a function of energy $E$, similarly to the spectral supercurrent, the total magnetization shows a monotonic decay with temperature.

\subsection{Triplet charge current} 
In the following, the case with a pure triplet charge current is explored and compared to the singlet charge current case. To reduce the number of parameters to vary, this discussion only considers the zero temperature case. We nevertheless expect a monotonic decay of the induced magnetization as the temperature approaches $T_c$.

To create a pure triplet charge current, as discussed in section \ref{sc:analytical_study}, the interface magnetizations have to be parallel. Moreover, we also have to set either the polarization or the spin-mixing angles to zero. Otherwise, a spin supercurrent will also flow through the junction \cite{gomperud_prb_15}. \st{Both these situations are here explored.}

We first consider the $P=0$ case, with a finite spin-mixing angle $G_\phi \neq 0$. This means only considering spin-independent transmission amplitudes from the proximate superconductor but with spin-dependent reflection terms.

\begin{figure}
    \centering
    \includegraphics[width = 0.5 \textwidth]{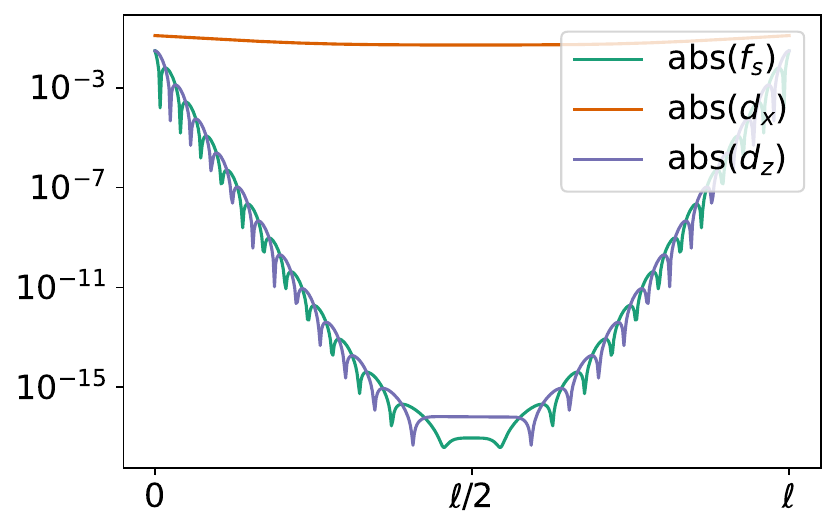}
    \caption{(Color online) The absolute value of the triplet components evaluated for $E/\Delta_0 = 0.14$. The figure shows that $f_s$ and $d_z$ are many orders of magnitude smaller than the long-ranged $d_x$ component.}
    \label{fig:components_in_L}
\end{figure}

We consider a situation where the $f_s$ component is negligible compared to the triplets in (L). This is achieved via the exchange field profile discussed in section \ref{sc:setup}. 
Fig. \ref{fig:components_in_L} show that both $f_s$ and $d_z$ die out over a short range into the material (L) when $\vecb{m}_1 = \vecb{m}_2 = (m,0,0)$ and $\vecb{h}(z) = (0,0,h(z))$.
The singlet and the $d_z$ triplet oscillate rapidly and decay quickly \cite{bergeret_rmp_05, eschrig_rpp_15}. In the middle region, $d_z$ and $f_s$ are many orders of magnitude smaller than $d_x$ and we conclude that the results from this section are pure triplet effects. 

\begin{figure}
    \centering
    \includegraphics[width = 0.5 \textwidth]{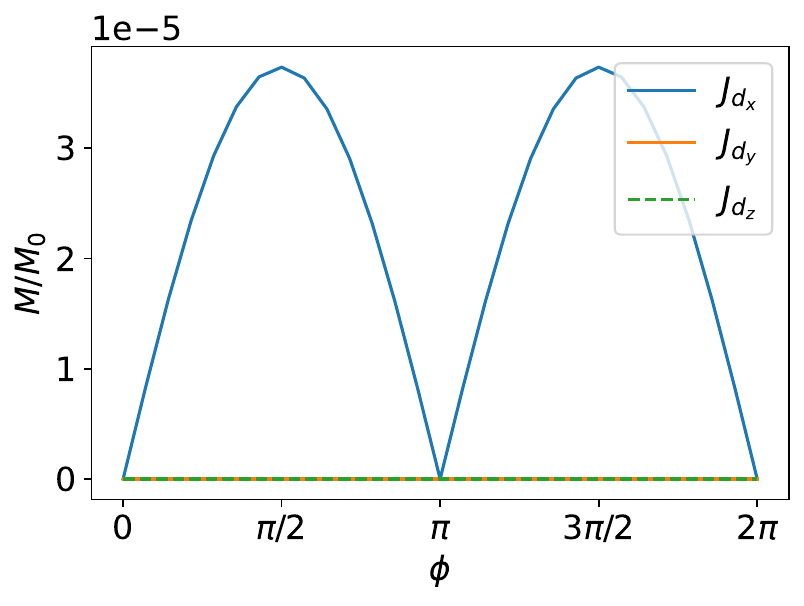}
    \caption{(Color online) The absolute value of the magnetization in material (R) induced by $d_x$, $d_y$, and $d_z$ carried charge current. It is seen that it is only the $d_x$ carried charge current that induces a magnetization. As seen, for certain triplet pair polarizations, a supercurrent ($J_{d_y}$ and $J_{d_z}$) does not induce magnetization in any direction.}
\label{fig:magnetization_from_Jdx_Jdy_Jdz}
\end{figure}

The triplet charge current can be divided into three components, $J_{d_x}, J_{d_y}$ and $J_{d_z}$, which from Sec. \ref{sc:analytical_study} are expected to give different results. We explore all these currents here. 
A $J_{d_x}$ current is created by using $\vecb{m}_1 = \vecb{m}_2 = (m,0,0)$ and $\vecb{h} = (0,0,h(z))$. 
In the same manner, a $J_{d_y}$ current is created by using $\vecb{m}_1 = \vecb{m}_2 = (0,m,0)$ and $\vecb{h} = (0,0,h(z))$, and a $J_{d_z}$ current is created with $\vecb{m}_1 = \vecb{m}_2 = (0,0,m)$ and $\vecb{h} = (h(z),0,0)$. 
The absolute value of the induced magnetization in (R) is plotted for the three different triplet cases as a function of $\phi$ in Fig. \ref{fig:magnetization_from_Jdx_Jdy_Jdz}. 
Consistent with the analysis in Sec. \ref{sc:analytical_study} only the $d_x$ carried current induces a magnetization in (R), 
which confirms the analytical results. 
The induced magnetization can thereby be utilized to distinguish $J_{d_x}$ from the other triplet currents. In other words, the induced spin-signal is strongly dependent on the polarization of the triplet Cooper pair supercurrent.

\begin{figure}[t!]
    \centering
        \includegraphics[width = 0.45\textwidth]{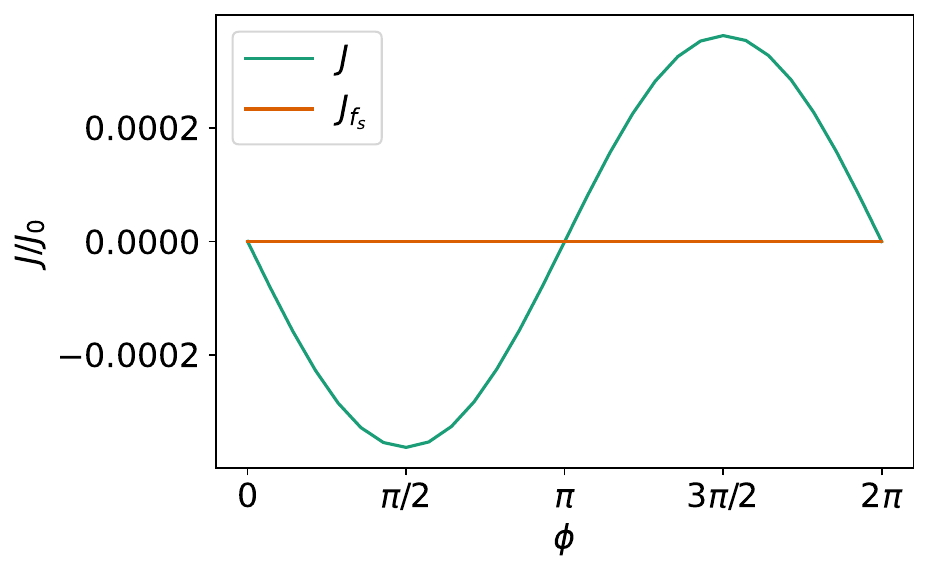}
        \caption{(Color online) Charge current as a function of $\phi$ in material (L). The green line in (a) shows the total current and the orange shows the singlet current. It is clear that the singlet current is negligible compared to the triplet current. The magnitude of the triplet charge current is invariant when changing $\alpha$.}
        \label{fig:triplet_charge_current_current_Pzero}
\end{figure}

\begin{figure}[t!]
    \centering
        \includegraphics[width = 0.45\textwidth]{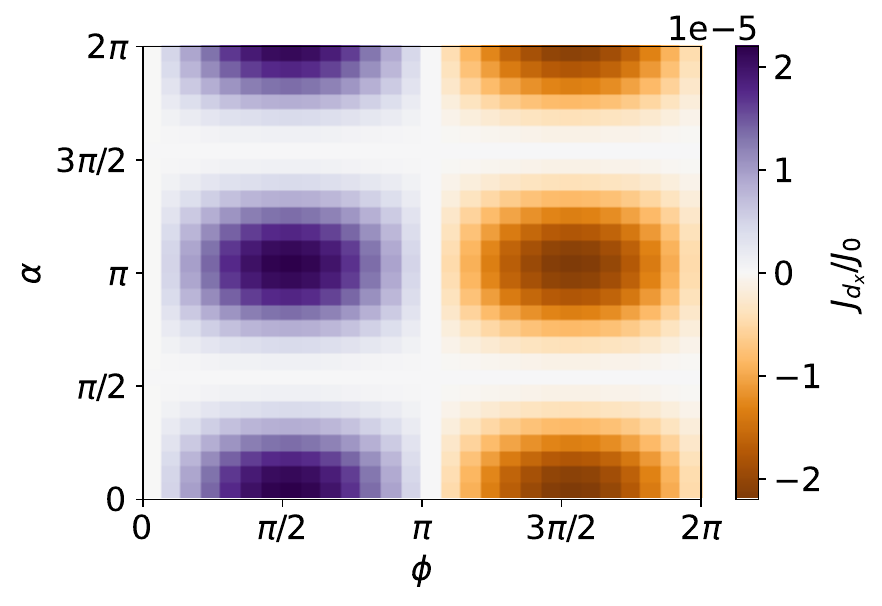}
        \caption{(Color online) The supercurrent carried by $d_x$ triplet Cooper pairs, which induce a magnetization in (R) through the spin-orbit coupled interface, as both the phase difference $\phi$ and interface angles in the  $xy$-plane, $\alpha$ is varied.}
        \label{subfig:tripletcurrent_Jdx}
\end{figure}

A triplet current does however not have to be carried by a pure $d_x,d_y$ or $d_z$, but could just as well be carried by any combination of $\vecb{d}$-triplet components. 
Therefore, rotation of the interface magnetization in $xy$-plane is investigated.
First, we consider the behavior of the charge current in material (L), shown in Fig. \ref{fig:triplet_charge_current_current_Pzero} as a function of the phase difference $\phi$. The current shows a standard current-phase relation and is independent on the value of $\alpha$. 
As seen in Sec. \ref{sc:analytical_study} the $G_\phi$-term creates a $\vecb{d}$-vector proportional to the interface magnetization. 
At $\alpha = 0$ the interface magnetizations are $\vecb{m}_1 = \vecb{m}_2 = (m,0,0)$, so it is natural that the charge current then will be carried only by the $d_x$ triplet. As $\alpha$ changes, the current will gradually be carried more and more by the $d_y$ component, until it at $\alpha = \pi/2$ is carried only by the $d_y$ components. 
Thus, as $\alpha$ is varied, the total current stays unchanged, however, $J_{d_x}$ does change, as shown in Fig. \ref{subfig:tripletcurrent_Jdx}.

\begin{figure}
    \centering
        \includegraphics[width = 0.45\textwidth]{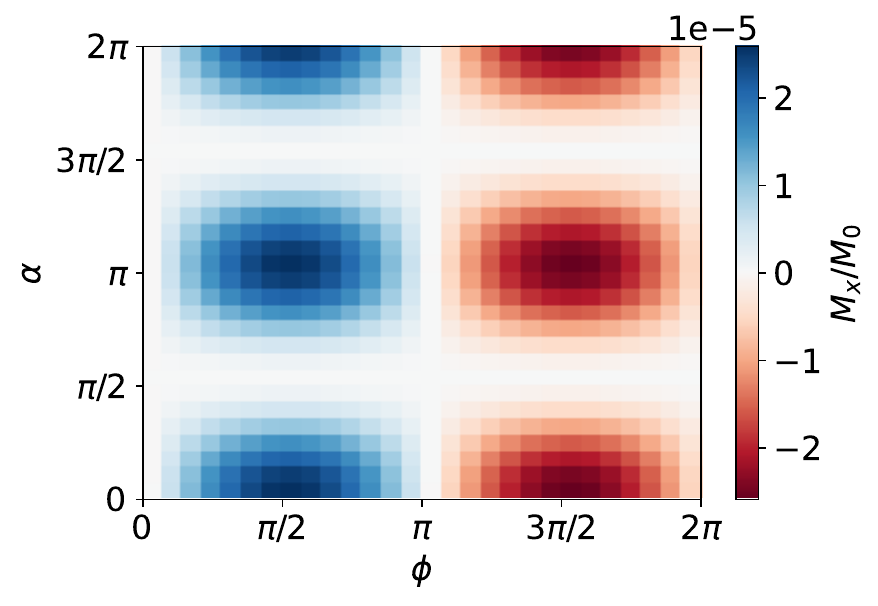}
        \caption{(Color online) Induced magnetization $m_x$ in (L) in the triplet supercurrent case where $G_\phi \neq 0, P= 0$. }
        \label{fig:mx_my_from_triplet_charge_current_mx}
\end{figure}

\begin{figure}
        \includegraphics[width = 0.45\textwidth]{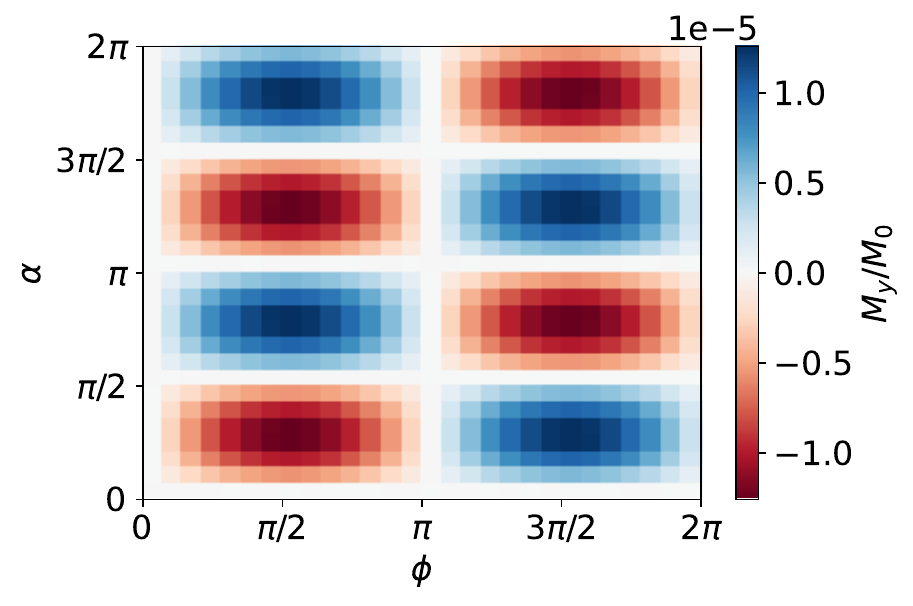}
        \caption{(Color online) Induced magnetization $m_y$ in (L) in the triplet supercurrent case where $G_\phi \neq 0, P= 0$. }
        \label{fig:mx_my_from_triplet_charge_current_my}
\end{figure}

The induced magnetization in material (R) in the triplet current case is plotted in Figs. \ref{fig:mx_my_from_triplet_charge_current_mx} and \ref{fig:mx_my_from_triplet_charge_current_my}.
As discussed in Sec. \ref{sc:analytical_study}, there exists a clear relation between $J_{d_x}$ and $m_{x,R}$. Fig. \ref{fig:mx_my_from_triplet_charge_current_mx} confirms this relation as the induced $m_x$ has the same form as $J_{d_x}$ from Fig. \ref{subfig:tripletcurrent_Jdx}.

The magnetization plots also show differences from the singlet current case. 
First of all, the induced $m_x$ varies with $\alpha$. In the singlet case there is no interface magnetization, however as varying $\alpha$ in the triplet case is the same as rotating the material (R) around material (L), it still makes sense to compare the $\alpha$ dependency of the singlet and the triplet case. 
What this means is that in the singlet case one would measure the same induced magnetization [in axes relative to (R)], no matter in what direction (R) was connected to (L). However, in the triplet case, the angle, in which (R) is connected to (L), makes a difference.

Besides the $\alpha$ dependency in $m_x$, an additional new signature arises in the triplet case, namely a magnetization component in the $y$-direction seen in Fig. \ref{fig:mx_my_from_triplet_charge_current_my}. 
In the singlet current case, only $m_x$ was induced, and thus the induced $m_y$ contributes to making the triplet and singlet currents distinguishable. 
From the figure it is seen that $m_y$ has a different $\alpha$ dependency than the $m_x$.  
This can be explained by realizing that we expect $\dxL$ and thus $\delz \dxL$ to have a $\cos(\alpha)$ dependency. This would according to the linearized spin-orbit coupled boundary conditions give $\dxR$ and $\fsR$ both a $\cos(\alpha)$ dependency. Thus, the induced magnetization $m_x$ in (R), which is a product of $\dxR$ and $\fsR$, naturally acquires a $\cos^{2} (\alpha)$ dependency as in the figure. 
Furthermore, $\dyL$ and thus also $\dyR$ we should expect to have a $\sin(\alpha)$ dependency. The magnetization in $y$-direction, a product of $\fsR$ and $\dyR$, should therefore, consistent with the figure, have a $\sin(\alpha)\cos(\alpha)$ dependency.

If we compare the singlet current case to the triplet current at $\alpha = 0$ the induced magnetization in (R) looks the same. 
However, by either growing the normal metal on the (L) region at different crystallographic orientations, or alternatively growing two normal metals to (L), both through a spin-orbit coupling interface, but at different angles, the induced magnetization would change strongly in the triplet case. 

\begin{figure*}[t!]
\includegraphics[width=1 \textwidth]{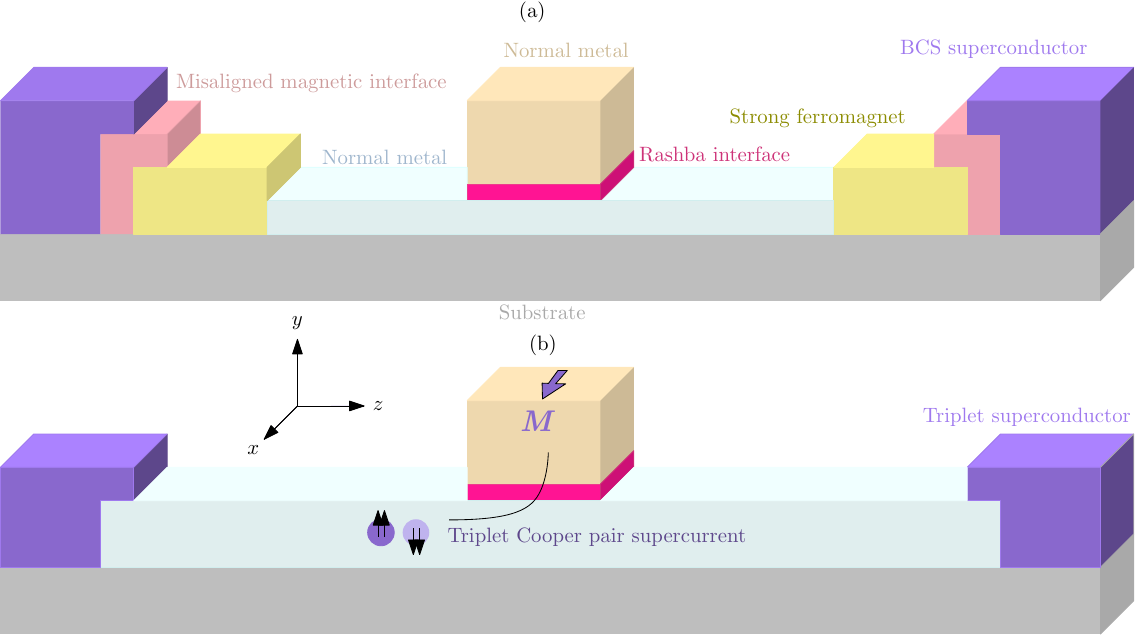}
	\caption{(Color online) Two possible experimental realizations of the proposed system. In (a), thin magnetic insulators with a magnetization in the $xy$-plane couple BCS superconductors to a strong ferromagnet which is polarized in the $z$-direction.  The strong ferromagnet filters out all superconducting correlations except equal-spin Cooper pairs along the magnetization direction. Adjusting the magnetization of the magnetic interfaces allows one to vary the spin-polarization of the triplet Cooper pairs carrying a supercurrent in the system. A simpler setup is shown in (b) where one uses intrinsic triplet superconductors. By growing these materials along different crystallographic axes relative the normal metal in the middle, the polarization of the triplet pairs, quantified by the $\boldsymbol{d}$-vector is varied. The drawback of setup (b) is that intrinsic $p$-wave triplet superconductivity is rare and only well-established in uranium-based compounds.}
	\label{fig:detailedsetup}
\end{figure*}

Similarly, as in the singlet case, we have checked that the disappearance of the magnetization at $\phi = 0$ and at $\phi = \pi$ can be explained by the symmetry properties under the $\tilde{\ldots}$ operation of the components. Thus, we do not expect that a spatial gradient in the triplet correlations is sufficient to induce a magnetization: only when a supercurrent is flowing, thus providing the triplet correlations with the correct $\tilde{\ldots}$-symmetry, is a magnetization induced.

To check the robustness of the results another parameter set was also investigated. The interface magnetizations are kept parallel, however, the polarization is turned on, adding the magnetoresistive and the $G_1$ term. Instead the spin-mixing is turned off, $G_\phi = 0$. From section \ref{sc:analytical_study} we see that the $G_{\text{MR}}$ term creates a triplet $\vecb{d}$ orthogonal to the interface magnetization and to $\vecb{h}$. 
In this case, a $\pi/2$ shift in $\alpha$ arises compared to the $P=0$ situation because of the induced triplet being orthogonal to the interface magnetization instead of parallel. Other than that, the form of the magnetization is similar, and the conclusions from the $P=0$ case hold.

\section{Conclusion}

In summary, we predict an experimental signature of current-carrying
triplet Cooper pairs in the form of an induced spin-signal. We show that a supercurrent carried only by triplet
Cooper pairs induces a non-local magnetization that is controlled by the polarization direction of the triplet
Cooper pairs. The dependence of the non-local magnetization on the polarization direction of the triplet pairs can be experimentally tested \textit{in situ}. Specifically, the component of the $\vecd$-vector carrying the supercurrent is determined by the magnetizations in Fig. \ref{fig:detailedsetup}. This provides a measurement protocol to directly use the spin-polarization of the triplet Cooper
pairs in supercurrents to transfer spin information in a dissipationless manner.

\text{ }\\

\begin{center}
    \textbf{ACKNOWLEDGMENTS}
\end{center}
 This work was supported by the Research
Council of Norway through Grant No. 323766 and its Centres
of Excellence funding scheme Grant No. 262633 “QuSpin.” Support from
Sigma2 - the National Infrastructure for High Performance
Computing and Data Storage in Norway, project NN9577K, is acknowledged.

 \appendix

\begin{widetext}
\section{Riccati parametrization of the SOC boundary conditions}

We now derive the Riccati parametrization of spin-orbit coupling boundary conditions given in Eq. \eqref{eq:SOC_boundary_conditions}. These boundary conditions take into account charge-spin conversion at the interface \cite{linder_prb_22}. The starting point is to perform the same operation on the right-hand side as we did to the left-hand side in Eq. \eqref{eq:left_hand_side_bc_riccati}
The first term is simply the Kuprianov-Lukichev boundary term \cite{kupriyanov_zetf_88}. The rest of the terms we go through one by one. For concreteness, the derivation below is done for a boundary condition  which has $g_R\partial_y g_R$ on the left-hand side of the boundary condition. 

The first term is $-\frac{2}{3}T_1^{2} p_F^{2} \com{\ghat_R}{\sigmaparvec\ghat_L\sigmaparvec}$, for which define the matrix 
\newcommand{\gL}[0]{g_L}
\newcommand{\tgL}[0]{\Tilde{g}_L}
\newcommand{\fL}[0]{f_L}
\newcommand{\tfL}[0]{\Tilde{f}_L}
\newcommand{\sigmax}[0]{\tau_x}
\newcommand{\sigmay}[0]{\tau_y}
\newcommand{\gmL}[0]{\gamma_L}
\newcommand{\NL}[0]{N_L}
\newcommand{\tNL}[0]{\tilde{N}_L}
\newcommand{\tgmL}[0]{\Tilde{\gamma}_L}
\begin{equation}
\Ubar^{(1)} = \sigmaparvec\ghat_L\sigmaparvec
= 
\begin{pmatrix}
\sigmax \gL \sigmax & - \sigmax \fL \sigmax^{*} \\
\sigmax^{*} \tfL \sigmax & - \sigmax^{*} \tgL \sigmax^{*} 
\end{pmatrix}
+ 
\begin{pmatrix}
\sigmaz \gL \sigmaz & - \sigmaz \fL \sigmaz^{*} \\
\sigmaz^{*} \tfL \sigmaz & - \sigmaz^{*} \tgL \sigmaz^{*} 
\end{pmatrix}.
\end{equation}

Thus the contribution from this term will according to Eq. \ref{eq:general_bc_term_riccati} be 
\begin{equation}
\begin{aligned}
&\frac{1}{2} N_R^{-1} \big ( \com{\ghat_R}{\sigmaparvec\ghat_L\sigmaparvec}^{1,2} - \com{\ghat_R}{\sigmaparvec\ghat_L\sigmaparvec}^{1,1} \gamma_R \big )
\\=&
\big[
    - (\sigmax \gL \sigmax + \sigmaz \gL \sigmaz) \gmR 
    - \sigmax \fL \sigmax^{*}  - \sigmaz\fL \sigma^{*}
    - \gmR ( \sigmax^{*} \tfL \sigmax + \sigmaz^{*} \tfL \sigmaz ) \gmR 
    \\&+ \gmR (- \sigmax^{*} \tgL \sigmax^{*} - \sigmaz^{*} \tgL \sigmaz^{*} )
\big ],
\end{aligned}
\end{equation}
and inserting $f = 2N\gamma$ and $g = (2N -1)$, we get
\begin{equation}
\begin{aligned}
     & -\sigmax \NL  \sigmax \gmR + \gmR - \sigmax \NL \gmL \sigmax^{*} - \gmR \sigmax^{*} 
    \tNL \tgmL \sigmax \gmR - \gmR \sigmax^{*} \tNL \sigmax^{*} 
    \\& -\sigmaz \NL  \sigmaz \gmR + \gmR - \sigmaz \NL \gmL \sigmaz^{*} - \gmR \sigmaz^{*} \tNL \tgmL \sigmaz \gmR - \gmR \sigmaz^{*} \tNL \sigmaz^{*} 
    .
\end{aligned}
\end{equation}

From the second term we define 
\begin{equation}
\begin{aligned}
    \Ubar^{(2)} =& \anticom{\sigmaparx}{\ghat_L \partial_y \ghat_L }
    \\=& 
    \begin{pmatrix}
        \sigmax [ \gL \dgL - \fL \dtfL ]  &
        \sigmax [ \gL \dfL - \fL \dtgL ]  \\
        - \sigmax^{*}[\tfL \dgL + \tgL \dtfL] &
        - \sigmax [- \tfL \dfL + \tgL \dtgL] 
    \end{pmatrix}
    \\&+
    \begin{pmatrix}
         [\gL \dgL - \fL \dtfL] \sigmax & - [\gL \dfL - \fL \dtgL ] \sigmax^{*} \\
         [\tfL \dgL + \tgL \dtfL] \sigmax  & - [- \tfL \dfL + \tgL \dtgL]\sigmax^{*}
    \end{pmatrix},
\end{aligned}
\end{equation}
which gives us the contribution to the right-hand side
\begin{equation}
\begin{aligned}
    \big [
    &-(\sigmax [ \gL \dgL - \fL \dtfL ] + [\gL \dgL - \fL \dtfL] \sigmax) \gmR 
    \\&+ \sigmax [ \gL \dfL - \fL \dtgL ] - [ \gL \dfL - \fL \dtgL ] \sigmax^{*} 
    \\&- \gmR ( - \sigmax^{*}[\tfL \dgL + \tgL \dtfL] + [\tfL \dgL + \tgL \dtfL] \sigmax  ) \gmR
    \\& \gmR (- \sigmax [- \tfL \dfL + \tgL \dtgL] - [- \tfL \dfL + \tgL \dtgL]\sigmax^{*})
    \big ].
\end{aligned}
\end{equation}

Using Eqs. \eqref{eq:riccati_derivative} and \eqref{eq:riccati_derivative_tilde} we see that we can write the following
\begin{equation}
\begin{aligned}
    &[ \gL \dgL - \fL \dtfL ] =  \NL [ \gmL' \tgmL - \gmL \tgmL' ] \NL, \\
    &[ \gL \dfL - \fL \dtgL ] =  \NL [\gmL' - \gmL \tgmL' \gmL ] \tNL.
\end{aligned}
\end{equation}

Thus the contribution to the right-hand side of the Riccati parametrized boundary conditions can be written as 
\begin{equation}
\begin{aligned}
    &\sigmax 2 \NL [\gmL' - \gmL \tgmL' \gmL ] \tNL - 2\NL [\gmL' - \gmL \tgmL' \gmL ] \tNL \sigmaxc
    \\&- \gmR \sigmaxc 2 \tNL [ \tgmL' \gmL - \tgmL \gmL'] \tNL - \gmR 2 \tNL [ \tgmL' \gmL - \tgmL \gmL'] \tNL \sigmaxc
    \\&- \sigmax 2 \NL [\gmL' \tgmL - \gmL \tgmL] \NL \gmR - 2 \NL [\gmL' \tgmL - \gmL \tgmL] \NL \sigmax \gmR 
    \\&+ \gmR  \sigmaxc 2 \tNL [\tgmL' - \tgmL \gmL' \tgmL ] \NL \gmR - \gmR 2 \tNL [\tgmL' - \tgmL \gmL' \tgmL ] \NL \sigmax \gmR. 
\end{aligned}
\end{equation}

The fourth term and fifth term  in Eq. \eqref{eq:SOC_boundary_conditions} can not be written as $\com{\gR}{\hat{U}}$, 
so we have to treat them differently. The two terms do, however, have the same form as each other. Therefore we only have to do the calculation once by performing the parametrization procedure on 
\newcommand{\rhohati}[0]{\hat{\rho}_i}
\begin{equation}
    \com{\rhohati}{\gRhat \rhohati \gRhat }.
\end{equation}

We write out 
\newcommand{\sigmai}[0]{\tau_i}
\newcommand{\sigmaic}[0]{\tau_i^{*}}
\begin{equation}
    \gR \rhohati \gR  = 
    \begin{pmatrix}
    \gR \sigmai \gR + \fR \sigmaic \tfR &   \gR \sigmai \fR + \fR \sigmaic \tgR \\
    -\tfR \sigmai \gR - \tgR \sigmaic \tfR  & -\tfR \sigmai \fR - \tgR \sigmaic \tgR 
    \end{pmatrix}.
\end{equation}

The upper left component of the whole expression then reads
\begin{equation}
    \com{\rhohati}{\gRhat \rhohati \gRhat}^{(1,1)}  = 
    \sigmai \gR \sigmai \gR + \sigmai \fR \sigmaic \tfR 
    - \gR \sigmai \gR \sigmai - \fR \sigmaic \tfR \sigmai, 
\end{equation}
and the upper right part reads 
\begin{equation}
    \com{\rhohati}{\gRhat \rhohati \gRhat}^{(1,2)}   = 
    \sigmai \gR \sigmai \fR + \sigmai \fR \sigmaic \tgR 
    +\gR \sigmai \fR \sigmaic + \fR \sigmaic \tgR \sigmaic.
\end{equation}

We write $g$ and $f$ in terms of the Riccati parametrized expressions and get
that the contribution from these terms to the right-hand side $\tfrac{1}{2}\NL^{-1} ( (1,2)- (1,1))$ $\tfrac{1}{2}\NR^{-1} ( (1,2)- (1,1))$ is

\begin{equation}
\begin{aligned}
    &\tfrac{1}{2} \NR^{-1} \big[\sigmai(2 \NR -1 ) \sigmai 2 \NR \gmR + \sigmai 2 \NR \gmR \sigmaic (2 \tNR - 1) 
    + (2\NR -1) \sigmai 2 \NR \gmR \sigmaic 
    \\& + 2 \NR \gmR \sigmaic (2\tNR - 1) \sigmaic 
    - [ \sigmai(2 \NR -1 ) \sigmai (2\NR -1 ) + \sigmai 2 \NR \gmR \sigmaic 2 \tNR \tgmR
    \\&- (2 \NR -1) \sigmai (2 \NR -1) \sigmai - 2 \NR \gmR \sigmaic 2 \tNR \tgmR
    ]\gmR \big ]
    \\&=
    -2 \gmR 
    + 2 \sigmai \NR \gmR \sigmaic 
    + 2 \gmR \sigmaic \tNR \sigmaic 
    + 2 \sigmai \NR \sigmai \gmR 
    + 2 \gmR \sigmaic \tNR \tgmR \sigmai \gmR. 
\end{aligned}
\end{equation}

Putting all of the terms together we get 
\begin{equation}
\begin{aligned}
    \dely \gmR  = 
    2 \tfrac{T_0^{2}}{D}  & ( 1 - \gmR \tgmL) \NL ( \gmR - \gmL ) 
    \\& - 2 \tfrac{2}{3} \tfrac{T_1^{2} p_F^{2}}{D}  \big ( -\sigmax \NL  \sigmax \gmR + \gmR - \sigmax \NL \gmL \sigmax^{*} - \gmR \sigmax^{*} \tNL \tgmL \sigmax \gmR - \gmR \sigmax^{*} \tNL \sigmax^{*} 
    \\& -\sigmaz \NL  \sigmaz \gmR + \gmR - \sigmaz \NL \gmL \sigmaz^{*} - \gmR \sigmaz^{*} \tNL \tgmL \sigmaz \gmR - \gmR \sigmaz^{*} \tNL \sigmaz^{*} 
     \big )
    \\& - m T_1T_0 \big ( +\sigmax 2 \NL [\gmL' - \gmL \tgmL' \gmL ] \tNL - 2\NL [\gmL' - \gmL \tgmL' \gmL ] \tNL \sigmaxc
    \\&- \gmR \sigmaxc 2 \tNL [ \tgmL' \gmL - \tgmL \gmL'] \tNL - \gmR 2 \tNL [ \tgmL' \gmL - \tgmL \gmL'] \tNL \sigmaxc
    \\&- \sigmax 2 \NL [\gmL' \tgmL - \gmL \tgmL] \NL \gmR - 2 \NL [\gmL' \tgmL - \gmL \tgmL] \NL \sigmax \gmR 
    \\&+ \gmR  \sigmaxc 2 \tNL [\tgmL' - \tgmL \gmL' \tgmL ] \NL \gmR - \gmR 2 \tNL [\tgmL' - \tgmL \gmL' \tgmL ] \NL \sigmax \gmR  \big) 
    \\&d\alpha^{2} \big(-2 \gmR 
    + 2 \sigmax \NR \gmR \sigmaxc 
    + 2 \gmR \sigmaxc \tNR \sigmaxc 
    + 2 \sigmax \NR \sigmax \gmR 
    + 2 \gmR \sigmaxc \tNR \tgmR \sigmax \gmR \big)
    \\&d\alpha^{2} \big(-2 \gmR 
    + 2 \sigmaz \NR \gmR \sigmazc 
    + 2 \gmR \sigmazc \tNR \sigmazc 
    + 2 \sigmaz \NR \sigmay \gmR 
    + 2 \gmR \sigmazc \tNR \tgmR \sigmaz \gmR \big).
\end{aligned}
\end{equation}

\section{Riccati parametrization of the spin active boundary conditions}

We now find the Riccati parameterized boundary conditions for the spin active interfaces given in Equation \eqref{eq:spin_active_bc}. 
All the terms on this boundary condition are on the same form as in Eq. \ref{eq:general_bc_term}. Just like for the spin-orbit coupling boundary conditions, the first term is simply the Kuprianov-Lukichev term, the rest we go through one term at a time starting with the $G_1$ term. Here we define

\newcommand{\mvec}[0]{\vecb{m}}

\begin{equation}
    U_{1} = \mhat \gRhat \mhat 
    = \begin{pmatrix}
        \mvec \cdot \sigmavec \gR \mvec \cdot \sigmavec & \mvec \cdot \sigmavec \fR \mvec \cdot \sigmavec^{*}   \\ 
        - \mvec \cdot \sigmavec^{*} \tfR  \mvec \cdot \sigmavec & - \mvec \cdot \sigmavec^{*} \tgR \mvec \cdot \sigmavec^{*} 
    \end{pmatrix},
\end{equation}

which gives the contribution to the right-hand side 
\begin{equation}
    -  \mvec \cdot\sigmavec \gR \mvec \sigmavec \gmL 
    +  \mvec \sigmavec \fR \mvec \sigmavec^{*} 
    + \gmL \mvec \sigmavec^{*} \tfR  \mvec \sigmavec \gmL 
    - \gmL \mvec \sigmavec^{*} \tgR \mvec \sigmavec^{*}. 
\end{equation}

For the second term we define the $U$-matrix
\begin{equation}
    U_{\text{MR}} = \anticom{\gRhat}{\mhat}  
    =
    \begin{pmatrix}
        \gR \mvec \cdot \sigmavec + \mvec \cdot \sigmavec \gR 
        & \fR \mvec \cdot \sigmavec^{*} + \mvec \cdot \sigmavec \fR
        \\ - \tfR \mvec \cdot \sigmavec - \mvec \cdot \sigmavec^{*} \tfR 
        & - \tgR \mvec \cdot \sigmavec^{*} - \mvec \cdot  \sigmavec^{*} \tgR 
    \end{pmatrix},
\end{equation}

which gives the contribution to the right-hand side

\begin{equation}
\begin{aligned}
    &- ( \gR \mvec \cdot \sigmavec + \mvec \cdot \sigmavec \gR  ) \gmL 
    \\&+ \fR \mvec \cdot \sigmavec^{*} + \mvec \cdot \sigmavec \fR
    \\&- \gmL (- \tfR \mvec \cdot \sigmavec - \mvec \cdot \sigmavec^{*} \tfR ) \gmL 
    \\&+ \gmL ( - \tgR \mvec \cdot \sigmavec^{*} - \mvec \cdot  \sigmavec^{*} \tgR  ).
\end{aligned}
\end{equation}

For the third term the $U$-matrix is simply $U_{\phi} = \mhat$, which gives the contribution 
\begin{equation}
    - \mvec \cdot \sigmavec \gmL 
    + \gmL \mvec \cdot \sigmavec^{*}.
\end{equation}

The total Riccati parameterized spin-active boundary conditions thus read 

\begin{equation}
\begin{aligned}
    \delz \gmL =& G_0   ( 1 - \gmL \tgmR) \NR ( \gmR - \gmL ) 
    \\&+ G_1  \big (
    \mvec \cdot \sigmavec \NR \gmR \mvec \cdot \sigmavec^{*}
    -  \gmL \mvec \cdot \sigmavec^{*} \tNR  \mvec \sigmavec^{*} 
    + m^{2} \gmL 
    \\&- \mvec \cdot \sigmavec \NR \mvec \cdot \sigmavec \gmL 
    + \gmL \mvec  \cdot \sigmavec^{*} \tNR \tgm \mvec \cdot \sigmavec \gmL 
    \big ) 
    \\&+ G_{\text{MR}} \big (
     \NR \gmR \mvec \cdot \sigmavec^{*} + \mvec \cdot \sigmavec \NR \gmR 
    - \gmL [ \tNR \mvec \cdot \sigmavec^{*} + \mvec \cdot \sigmavec^{*} \tNR - \mvec \cdot \sigmavec^{*}  ]
    \\&- [\NR \mvec \cdot \sigmavec - \mvec \cdot \sigmavec + \mvec \cdot \sigmavec \NR] \gmL
    + \gmL [\tNR \tgmR \mvec \cdot \sigmavec + \mvec \cdot \sigmavec^{*} \tNR \tgmR ] \gmL \big )
    \\ &- i G_\phi  \big (
    - \mvec \cdot \sigmavec \gmL 
    + \gmL \mvec \cdot \sigmavec^{*} \big ).
\end{aligned}
\end{equation}

\end{widetext}

\section{Analysis of the symmetry of the anomalous Green function under the $\tilde{\ldots}$-operation}

\begin{figure}[H]
    \centering
         \includegraphics[width = 0.4\textwidth]{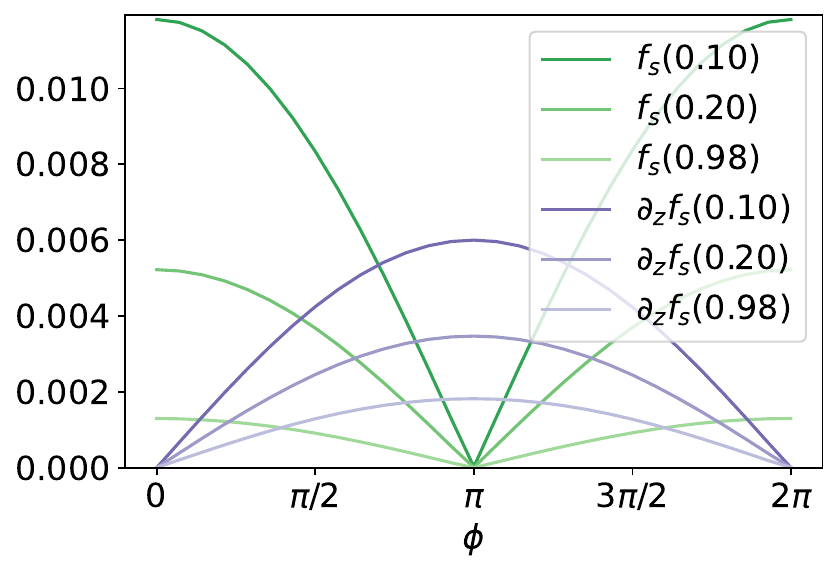}
         \caption{(Color online) The absolute value of the singlet component $f_s(E/\Delta_0)$ and its derivative in material (L) at $z_0 = l/2$.}
         \label{subfig:singlet_abs_fs_dfs_half}
\end{figure}

\begin{figure}[H]
        \includegraphics[width = 0.4\textwidth]{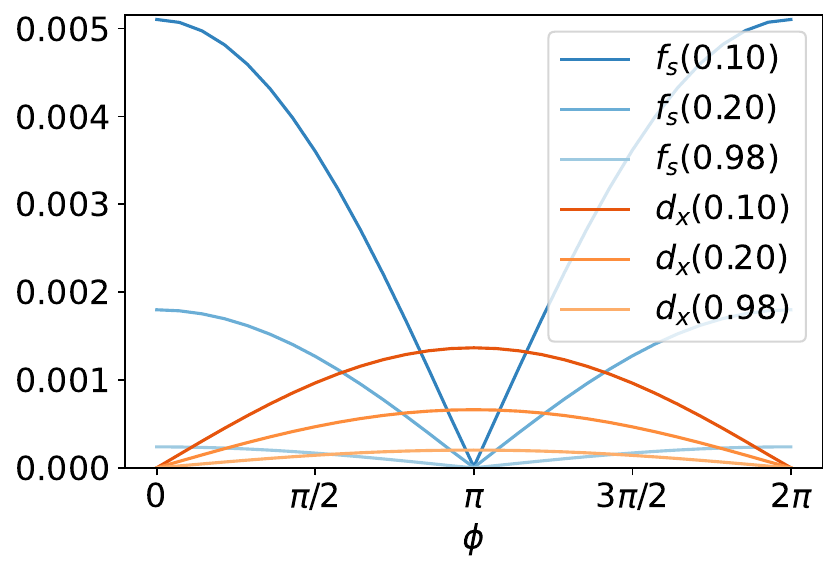}
         \caption{(Color online) The absolute value of the components $\fs(E/\Delta_0)$ and $\dx(E/\Delta_0)$ in material (R) as a function of $\phi$ for different energies, $E/ \Delta_0$. at $z_0 = l/2$.}
         \label{subfig:singlet_abs_fs_dx_half}
\end{figure}

We here analyze whether the disappearance of the magnetization at $J = 0$ is caused by the spatial gradient of the $\fsR$ and $\dxR$ components at $\phi = 0, \pi$ or by the symmetry properties under the $\tilde{\ldots}$ operation of the triplet and singlet component. 
To investigate this, consider Figs. \ref{subfig:singlet_abs_fs_dfs_half} and \ref{subfig:singlet_abs_fs_dx_half}. 
The first situation explored is when material (R) is connected to the middle of material (L), $z_0 = l/2$. It is seen from the figure that in this case the singlet component, $\fsL$, is zero at $\phi =\pi$ while the derivative, $\delz \fsL$, is zero at $\phi = 0$. The figure shows that this also causes $\dxR$ to vanish at $\phi = 0$ and $\fsR$ at $\phi = \pi$. Thus, a finite derivative of $f_s$ is not sufficient to induce a magnetization in (R).

\begin{figure}[H]
        \includegraphics[width = 0.4\textwidth]{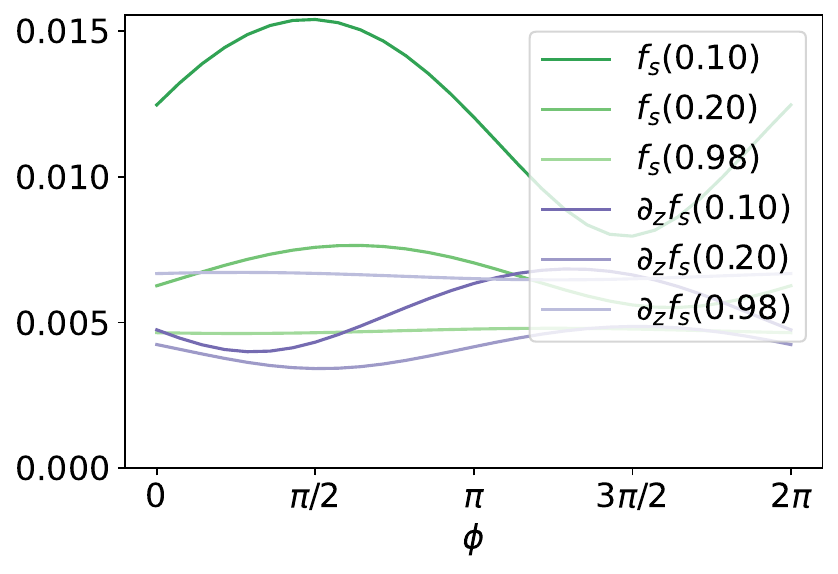}
         \caption{(Color online) The absolute value of the singlet component $f_s(E/\Delta_0)$ and its derivative in material (L) at $z_0 = l/4$}
         \label{subfig:singlet_abs_fs_dfs_quarter}
\end{figure}

\begin{figure}[H]
        \includegraphics[width = 0.4\textwidth]{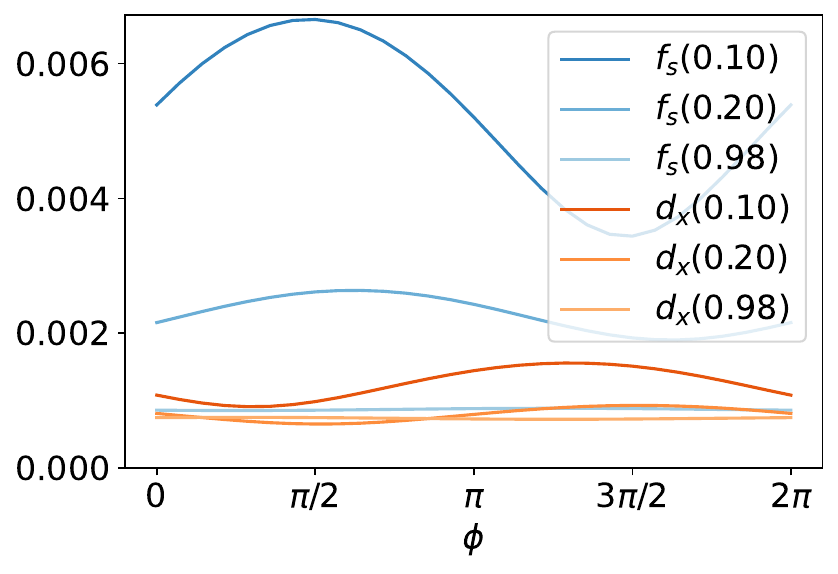}
         \caption{(Color online) The absolute value of the components $\fs(E/\Delta_0)$ and $\dx(E/\Delta_0)$ in material (R) as a function of $\phi$ for different energies, $E/\Delta_0$. at $z_0 = l/4$.}
         \label{subfig:singlet_abs_fs_dx_quarter}
\end{figure}

To check if it is the symmetry of the anomalous Green function under the $\tilde{\ldots}$-operation that dictates when a finite magnetization occurs, we considered a different situation where $z_0 = l/4$ so that the material (R) is no longer connected to the middle of (L). 
The result is shown in Figs. \ref{subfig:singlet_abs_fs_dfs_quarter} and \ref{subfig:singlet_abs_fs_dx_quarter}. 
It is here seen that neither $\fsL$, nor $\delz \fsL$ vanishes at any $\phi$. 
Therefore also $\dxR$ and $\fsR$ is finite at every $\phi$. 
It was checked that the magnetization looks exactly like Fig. \ref{subfig:singlet_Mx_My_Mz}, also for $z_0= l/4$. 
Thus we can conclude that it is the $\tilde{\ldots}$-operation symmetry, which is influenced by whether or not a supercurrent flows, that causes the magnetization to vanish at $\phi = 0, \pi$.

\bibliography{main}

\newpage

\text{ }\\

\end{document}